\documentclass[prr,aps,superscriptaddress,twocolumn,10pt]{revtex4-2}

\usepackage{tikz}
\usetikzlibrary{
    decorations.pathreplacing,
    decorations.pathmorphing,
    arrows
}

\usetikzlibrary{positioning, calc,intersections,shapes,matrix,fit}
\usepackage[normalem]{ulem} 
\usepackage{float}
\usepackage{graphicx}
\usepackage{dcolumn}
\usepackage{bm}
\usepackage{xcolor}
\usepackage{appendix}
\usepackage{caption}
\usepackage{subcaption}
\usepackage{multirow}
\usepackage{booktabs}
\usepackage{amsmath}
\usepackage{amsfonts}
\usepackage[braket, qm]{qcircuit}
\usepackage{url}
\usepackage{array}
\usepackage{titlesec}
\usepackage{etoolbox}

\usepackage{amsthm}

\theoremstyle{definition}
\usepackage[english]{babel}
\usepackage{algpseudocode}
\usepackage{algorithm}

\newcommand{\afchem}{Department of Chemistry, University of California, Berkeley, California 94720, USA}
\newcommand{\afbqic}{Berkeley Center for Quantum Information and Computation, Berkeley, California 94720, USA}
\newcommand{\ciqc}{Challenge Institute for Quantum Computation, Berkeley, California 94720, USA}
\newcommand{\afphys}{Department of Physics, University of California, Berkeley, California 94720, USA}

\begin{document}
\preprint{APS/123-QED}

\title{Towards Heisenberg Scaling: Measurement-Efficient Non-Orthogonal Quantum Eigensolver}

\author{Hang Ren}
\email{hangren@berkeley.edu}
\affiliation{\afbqic}
\affiliation{\afchem}
\affiliation{\ciqc}
\author{Yipei Zhang}\thanks{Equal contributions.}
\affiliation{\afbqic}
\affiliation{\afchem}
\affiliation{\ciqc}
\author{Thilo Scharnhorst}
\affiliation{\afbqic}
\affiliation{\afchem}
\affiliation{\ciqc}
\affiliation{\afphys}
\author{K. Birgitta Whaley}
\email{whaley@berkeley.edu}
\affiliation{\afbqic}
\affiliation{\afchem}
\affiliation{\ciqc}
\date{\today}

\begin{abstract}
The Non-Orthogonal Quantum Eigensolver (NOQE)~\cite{baek2023say}
provides an accurate framework for electronic-structure calculations, but the estimation of its Hamiltonian and overlap matrix elements relies on sampling and requires $\mathcal{O}(1/\varepsilon^2)$ circuit repetitions to achieve additive precision $\varepsilon$.
Here, we reformulate this matrix-element estimation step as a collection of amplitude-estimation tasks and integrate iterative quantum amplitude estimation into the NOQE workflow.
The resulting protocol achieves near-Heisenberg query complexity $\tilde{\mathcal{O}}(1/\varepsilon)$ for these estimation tasks, by replacing incoherent statistical averaging with coherent amplitude amplification.
We present explicit circuit constructions and the corresponding implementation procedure.
Numerical simulations for the electronic states of the hydrogen molecule show that the proposed method reaches chemical accuracy with substantially fewer total queries than the original sampling-based protocol. Overall, this work provides a measurement-efficient route to high-precision energy estimation and illustrates how sampling-limited quantum algorithms can be systematically reformulated to leverage quantum coherence and achieve lower measurement costs.
\end{abstract} 

\maketitle

\section{Introduction}

Quantum chemistry is a major application area for quantum computing, and many
quantum algorithms have been proposed for electronic structure calculations \cite{mcardle2020quantum,bauer2020quantum,motta2022emerging}.
Despite their differences in algorithmic structure, all such approaches
ultimately require measurements to extract physical quantities from a quantum
device. Although measurement has been developed as a resource for controlling quantum systems \cite{PRXQuantum.5.020366, zhang2025optimal, zhang2025solving}, its sample complexity remains a major bottleneck in algorithmic settings, and the results of the computation cannot be
accessed without this measurement step.

For near-term algorithms such as the variational quantum
eigensolver (VQE), observables are typically estimated by sampling Pauli operators \cite{cerezo2021variational,tilly2022variational}. This measurement process faces two well-known challenges. First, realistic molecular Hamiltonians can contain a
large number of Pauli terms. Second, and more fundamentally, estimating expectation values
through averaging the outcomes of $n$ repeated projective measurements yields a statistical error $\varepsilon$ that
scales as $\mathcal{O}(1/\sqrt{n})$. Consequently, achieving a target precision
$\varepsilon$ requires $n = \mathcal{O}(1/\varepsilon^2)$ independent measurements. Although
various techniques have been developed to reduce the number of measured Pauli
terms \cite{huggins2021efficient,patel2025quantum}, the sampling-limited
$\mathcal{O}(1/\varepsilon^2)$ scaling remains a fundamental constraint for expectation-value
estimation based on independent measurements.

Quantum phase estimation (QPE) \cite{kitaev1995quantum} provides a conceptually different route and can
achieve $\tilde{\mathcal{O}}(1/\varepsilon)$ scaling in the number of
queries to the underlying oracle. However,
QPE typically relies on deep, coherent circuits involving the quantum Fourier transform,
placing it beyond the capabilities of current noisy
intermediate-scale quantum hardware and making it more suitable for future
fault-tolerant quantum computers.

To address the optimization overhead of VQE while avoiding the circuit-depth
requirements of QPE, the non-orthogonal quantum eigensolver (NOQE)
\cite{baek2023say} was introduced as a non-variational approach to molecular
energy estimation.
Unlike VQE, NOQE avoids the iterative parameter optimization loop and the associated
measurement overhead, and can be implemented using relatively shallow quantum
circuits.
The method constructs a low-dimensional subspace spanned by multiple
non-orthogonal ansatz states and estimates molecular energies by solving a
generalized eigenvalue problem within this subspace.
The use of multiple ansatz states enables the description of strong
correlation effects, while unitary coupled-cluster dressing systematically
captures weak correlations.
As a result, NOQE avoids both variational optimization and the use of quantum
Fourier transforms, while retaining the ability to achieve accurate energy
estimation, with the additional benefit of providing both practical~\cite{baek2023say} and algorithmic~\cite{leimkuhler2025exponential} quantum advantage.

Despite these advantages, the practical cost of NOQE in its original form~\cite{baek2023say} is dominated by the
estimation of Hamiltonian and overlap matrix elements. These matrix elements are obtained from sampled quantum measurements, so achieving additive
precision $\varepsilon$ requires $\mathcal{O}(1/\varepsilon^2)$ repetitions. We have recently shown that developments such as shadow-tomography-based approaches
\cite{ren2026error} can reduce this cost in certain
regimes, but the applicability of such reductions remains limited to systems of relatively small size and high target precision.

This raises a natural question: can the sampling-limited matrix-element
estimation in NOQE be fundamentally improved while preserving its favorable
features and near-term implementability?

In this work, we answer this question in the affirmative by reformulating NOQE
within the framework of iterative quantum amplitude estimation (IQAE) \cite{grinko2021iterative}. By
mapping the estimation of Hamiltonian and overlap matrix elements to amplitude
estimation tasks, we replace incoherent statistical averaging with coherent
phase accumulation prior to measurement. This reduces the query complexity of these estimations
from $\mathcal{O}(1/\varepsilon^2)$ to $\tilde{\mathcal{O}}(1/\varepsilon)$, without invoking
quantum Fourier transforms. We refer to this
resulting method as IQAE--NOQE.

The main contributions of this work are as follows. First, we reformulate the
matrix-element estimation problem in NOQE as an amplitude estimation problem.
Second, we integrate iterative quantum amplitude estimation into the NOQE
framework and provide explicit circuit constructions and post-processing
procedures. More generally, 
the reformulation enables different amplitude-estimation variants to be
incorporated modularly, beyond the specific IQAE implementation studied here. Third, we analyze the resulting resource--accuracy trade-offs and
systematically compare IQAE--NOQE with the original sampling-based NOQE. Finally, we demonstrate the performance of the proposed approach through numerical simulations of molecular energy estimation, showing that accurate energies can be obtained with substantially fewer queries than in the original protocol, while keeping the circuit size accessible within the near-term or early fault-tolerant regime.

The remainder of this paper is organized as follows. Sections~\ref{sec:NOQE} and \ref{sec:IQAE}
review the NOQE framework and the basic principles of IQAE. Section
\ref{sec:IQAE-NOQE} introduces the IQAE--NOQE algorithm and its circuit
implementation. Section~\ref{sec:result} presents numerical results, and Section
\ref{sec:summary} discusses the implications of the proposed approach and outlines future directions.

\section{Overview of NOQE}
\label{sec:NOQE}

\begin{figure}[htbp]
    \centering
    \includegraphics[width=\linewidth]{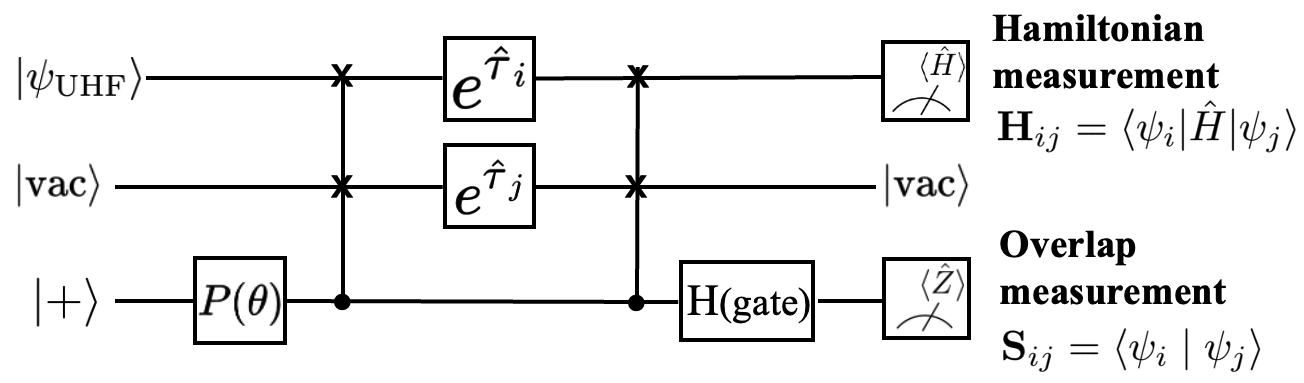}
    \caption{Circuit diagram of the original NOQE method~\cite{baek2023say}, modified to absorb the basis rotations into the ansatz operators $e^{\hat{\tau}_{i(j)}}$ (see text). Two $N$-qubit non-orthogonal ansatz states, $|\psi_i\rangle$ and $|\psi_j\rangle$, are prepared and combined using controlled operations to enable modified Hadamard tests that measure $H_{ij}$ and $S_{ij}$. After measuring these matrix elements, one solves the generalized eigenvalue problem (Eq.~\ref{eq:gep}) to obtain the electronic energies. This circuit requires $2N+1$ qubits. A more compact version using $N+1$ physical qubits and requiring longer circuits was also presented in Ref.~\cite{baek2023say}.}
    \label{fig:noqe circuit}
\end{figure}

In this section, we briefly review the non-orthogonal quantum eigensolver
(NOQE), focusing on the parts most relevant to matrix
element estimation and measurement cost.

The central idea of NOQE is to approximate molecular eigenstates within a
low-dimensional subspace of dimension $M$ that is spanned by a set of non-orthogonal ansatz states
$\{|\psi_i\rangle\}_{i=1}^M$. These ansatz states are typically generated from
unrestricted Hartree--Fock (UHF) \cite{pratt1956unrestricted} solutions that incorporate strong correlation and are further dressed by a unitary
coupled-cluster (UCC) ansatz to additionally incorporate correlation effects. 
\begin{equation}\label{eq: ucc+hf} |\psi_i\rangle = e^{\hat{\tau}_i}|\psi_{\mathrm{UHF}}\rangle. \end{equation}

Here \(e^{\hat{\tau}_i}\) denotes the full state-preparation unitary for the \(i\)th ansatz state. Since different UHF-based ansatz states may be defined in different single-particle bases, matrix elements must be evaluated after transforming the two states to a common basis. Following Ref.~\cite{baek2023say}, we absorb the required orbital-rotation unitaries, \( \hat{U}_{i\rightarrow 1} \) and \( \hat{U}_{j\rightarrow 1} \), into the corresponding ansatz circuit primitives \(e^{\hat{\tau}_i}\) and \(e^{\hat{\tau}_j}\), as indicated in Fig.~1.
The UCC ansatz parameters can be approximated through classical preprocessing
at polynomial cost, e.g., using second-order many-body perturbation theory \cite{baek2023say}.

NOQE constructs the Hamiltonian matrix
$H_{ij}=\langle \psi_i| H|\psi_j\rangle$ and the overlap matrix
$S_{ij}=\langle \psi_i|\psi_j\rangle$. Once these matrix elements are available,
the electronic energies are obtained by solving the generalized eigenvalue
problem
\begin{equation} \label{eq:gep}
\mathbf{H} \vec{c}=\mathbf{E S} \vec{c}
\end{equation}
on a classical computer. 

Figure~\ref{fig:noqe circuit} illustrates the circuit structure used to estimate
$H_{ij}$ and $S_{ij}$ in the original NOQE framework. Two non-orthogonal
ansatz states $|\psi_i\rangle$ and $|\psi_j\rangle$ are prepared on separate
registers and combined using a modified Hadamard test
to measure the overlap and
Hamiltonian matrix elements. In practice, the Hamiltonian expectation values are
estimated by measuring all terms in a Pauli operator decomposition of the electronic Hamiltonian. Ref.~\cite{baek2023say} also proposed an alternative NOQE circuit construction that differs from Fig.~\ref{fig:noqe circuit} by reducing the required qubit count from $2N+1$ to $N+1$ 
while increasing the circuit depth, with the overall query complexity remaining unchanged. 
For clarity, in this work we focus solely on the circuit shown in Fig.~\ref{fig:noqe circuit}.

While NOQE avoids the repeated measurements associated with variational
optimization loops, the estimation of matrix elements still relies on sampling
from projective measurements. Each measurement yields a binary outcome, and the
desired expectation values are obtained by statistical averaging over many
repetitions. As a result, the measurement cost of estimated matrix elements
scales as $\mathcal{O}(1/\varepsilon^2)$ with the target precision $\varepsilon$, making measurements a significant bottleneck in high-accuracy
or large-subspace NOQE calculations on near-term quantum hardware.

This sampling-limited measurement structure motivates the developments to improve this scaling by integrating an IQAE measurement protocol that we present in this work. In the following sections, we will introduce IQAE and then show how the matrix element
estimation step in NOQE can be reformulated within the IQAE framework to yield a substantial reduction in measurement
complexity, while still preserving the favorable features of the original NOQE approach.

\section{Iterative Quantum Amplitude Estimation}\label{sec:IQAE}

\begin{algorithm}[H]
\caption{Iterative Quantum Amplitude Estimation:
$\operatorname{IQAE}(\mathcal{A},\varepsilon,\delta)$.}
\label{alg:iqae}
\begin{algorithmic}[1]
\Require State-preparation unitary $\mathcal{A}$ acting on $n+1$ qubits, target accuracy $\varepsilon$, confidence level $1-\delta$
\Ensure Estimate $\hat{a}$ of $a=\sin^2(\theta)$ such that $|a-\hat{a}|\le \varepsilon$ with probability at least $1-\delta$

\State Construct the Grover operator $\mathcal{Q}=\mathcal{A}\mathcal{S}_{0}\mathcal{A}^\dagger \mathcal{S}_{\psi_0}$

\State Initialize iteration index $i \gets 0$
\State Initialize confidence interval $[\theta_l,\theta_u]\gets[0,\pi/2]$
\State Set the maximum number of iterations $T = \operatorname{log}(\pi/8\varepsilon)$
\State Set the per-iteration failure probability $\delta_i = \delta / T$

\While{$\theta_u-\theta_l>2\varepsilon$}
    \State $i \gets i+1$
    \State Choose $k_i$ such that the amplified interval $[(2k_i+1)\theta_l,(2k_i+1)\theta_u]$ lies entirely within a single quadrant
    \State Execute $N_i$ shots of the circuit $\mathcal{Q}^{k_i}\mathcal{A}\ket{0^{n+1}}$
    \State Compute the empirical frequency $\hat{p}_i$ of measuring outcome $1$ on the ancilla qubit (i.e., $\hat{p}_i \approx \sin^2((2k_i+1)\theta)$)
    \State Form a confidence interval $[p_i^{\min},p_i^{\max}]$ for $p_i$ using a chosen binomial-confidence method (e.g., Chernoff--Hoeffding or Clopper--Pearson)
    \State Map this interval to an updated interval $[\theta_l,\theta_u]$ using the monotonicity of $\sin^2(\cdot)$ within the chosen quadrant
\EndWhile

\State \Return $\hat{a} \leftarrow \sin^2\!\left(\frac{\theta_l+\theta_u}{2}\right)$
\end{algorithmic}
\end{algorithm}

In this section, we review the quantum amplitude estimation (QAE) problem and introduce the specific variant—Iterative Quantum Amplitude Estimation (IQAE), summarized in Algorithm~\ref{alg:iqae}, that will be integrated into our IQAE--NOQE protocol. The goal here is to summarize the essential elements of IQAE that are needed to develop our protocol; for full technical details of IQAE, see Ref.~\cite{grinko2021iterative}.

\subsection{The Amplitude Estimation Problem}

In the problem of quantum amplitude estimation, we are given a unitary operator $\mathcal{A}$ acting on the initial state $\ket{0^{n+1}}$ such that
\begin{equation}\label{eq:qae}
\mathcal{A}\ket{0^{n+1}}
=
\sin(\theta)\ket{\psi_1}\ket{1}
+
\cos(\theta)\ket{\psi_0}\ket{0},
\end{equation}

Here $\ket{\psi_0}$ and $\ket{\psi_1}$ denote generic normalized states on a system register of $n$ qubits associated with the two possible outcomes of the final $(n+1)$th readout qubit in the standard amplitude-estimation formulation~\cite{grinko2021iterative}. At this stage they are not specific to NOQE. The task is to estimate
\begin{equation}
a = \sin^2(\theta)
\end{equation}
up to additive error $\varepsilon$ with confidence at least $1-\delta$, i.e., to obtain an estimate $\hat{a}$ such that $|a-\hat{a}|\le \varepsilon$ with probability at least $1-\delta$.

The naive way to complete this task is to perform a projective measurement on the last qubit and empirically estimate the probability of obtaining outcome $1$. Since each measurement yields a Bernoulli random variable with mean $a$, achieving additive error $\varepsilon$ requires $\mathcal{O}(1/\varepsilon^2)$ repetitions of the circuit, with each shot querying $\mathcal{A}$ once. This corresponds to standard Monte Carlo scaling and is fundamentally sampling-limited.

\subsection{Quantum Amplitude Estimation}

Quantum amplitude estimation (QAE) is a family of quantum algorithms developed to achieve $\tilde{\mathcal{O}}(1/\varepsilon)$ query complexity with respect to the unitary $\mathcal{A}$. The central idea is to encode the amplitude $a$ into a phase $\theta$ and amplify it coherently before measurement.

Define the Grover operator
\begin{equation}
\mathcal{Q} = \mathcal{A}\mathcal{S}_{0}\mathcal{A}^{\dagger}\mathcal{S}_{\psi_0},
\end{equation}
where
\begin{equation}
\mathcal{S}_{\psi_0} = I - 2\ket{\psi_0}\bra{\psi_0}\otimes\ket{0}\bra{0},
\quad
\mathcal{S}_{0} = I - 2\ket{0^{n+1}}\bra{0^{n+1}}
\end{equation}
are reflection operators about the target state and the initial state $\ket{0^n}$, respectively. The reflection about the initial state can be effectively implemented by the reflection about the ancilla state $\ket{0}$, which allows oblivious QAE without access to the target state $|\psi_0\rangle$. The operator $\mathcal{Q}$ acts as a rotation in the two-dimensional subspace spanned by $\ket{\psi_1}\ket{1}$ and $\ket{\psi_0}\ket{0}$. In particular,
\begin{equation}
\mathcal{Q}^k \mathcal{A}\ket{0^{n+1}}
=
\sin\big((2k+1)\theta\big)\ket{\psi_1}\ket{1}
+
\cos\big((2k+1)\theta\big)\ket{\psi_0}\ket{0}.
\end{equation}
By applying $\mathcal{Q}$ with different powers $k$, one can extract information about $a$ (encoded in $\theta$) with a query complexity achieving Heisenberg scaling $\tilde{\mathcal{O}}(1/\varepsilon)$, improving on the standard quantum limit $\mathcal{O}(1/\varepsilon^2)$. We shall define the query complexity of QAE as the total number of applications of the unitary $\mathcal{A}$, which is twice the total number of applications of the Grover operator $\mathcal{Q}$.

The traditional QAE algorithm relies on quantum phase estimation (QPE), which uses multiple controlled-$\mathcal{Q}$ operations followed by an inverse quantum Fourier transform. While this achieves the desired Heisenberg scaling, it requires long coherent time and multi-qubit controlled operations. Such circuits are typically considered practical only in the fault-tolerant regime and are beyond the capability of current noisy hardware.

\subsection{IQAE Algorithm}

Recently, several variants of QAE have been proposed to avoid QPE and thus eliminate controlled $\mathcal{Q}^k$ operations and the quantum Fourier transform, while preserving (near) Heisenberg-limited query complexity~\cite{aaronson2020quantum,fukuzawa2023modified,callison2022improved,giurgica2022low}.

In these algorithms, the powers $\mathcal{Q}^k$ are not applied within a single large controlled circuit. Instead, different powers of $\mathcal{Q}$ are applied in sequential circuits. At each iteration, the power $k$ is either chosen according to a predefined schedule or adaptively based on previous measurement results. After each round of measurements, a classical confidence interval for $\theta$ is updated. This process is performed iteratively until the confidence interval is narrowed to the desired precision.

Among these QAE variants without QPE, we focus here on Iterative Quantum Amplitude Estimation (IQAE), which has been shown to offer the best practical performance to date~\cite{grinko2021iterative}. In particular, IQAE can be implemented with current hardware with a modest increase in the circuit depth while achieving favorable query complexity.

IQAE iteratively updates a confidence interval $[\theta_l, \theta_u]$ for $\theta$. The full procedure is summarized in Algorithm~\ref{alg:iqae}. Specifically, at iteration $i$, given the current interval $[\theta_l, \theta_u]$, the next power $k_{i+1}$ of $\mathcal{Q}$ is chosen such that the amplified interval
\begin{equation}
[(2k_{i+1}+1)\theta_l,\,(2k_{i+1}+1)\theta_u]
\end{equation}
lies entirely within an interval where $\sin^2(\cdot)$ is monotonic (i.e., within a single quadrant of $[0,2\pi)$). This condition ensures that the mapping between the measured probability
\begin{equation}
p_{k_{i+1}} = \sin^2\big((2k_{i+1}+1)\theta\big)
\end{equation}
and the underlying angle $\theta$ is unambiguous. As a result, the new confidence interval for $\theta$ can be uniquely determined from the measurement outcomes at that iteration.

This iterative procedure achieves near-Heisenberg scaling up to logarithmic factors. More precisely, IQAE estimates the amplitude $a$ up to additive error $\varepsilon$ with probability at least $1-\delta$ using 

\begin{equation}
\mathcal{O}\!\left(
\frac{1}{\varepsilon}
\log\!\left(
\frac{1}{\delta}
\log\!\frac{1}{\varepsilon}
\right)
\right)
\end{equation}

queries to $\mathcal{A}$ and $\mathcal{A}^\dagger$.

The above summary provides the basic background of IQAE. In the following section, we show how to reformulate the key matrix-element estimation step in NOQE as amplitude estimation tasks and to then integrate IQAE into the NOQE framework.

\section{IQAE--NOQE Algorithm}\label{sec:IQAE-NOQE}

In this section, we reformulate the matrix-element estimation step of NOQE as a set of
iterative quantum amplitude estimation (IQAE) tasks, yielding the IQAE--NOQE protocol. We will see that estimation of both overlap and Hamiltonian matrix elements reduces to estimating an amplitude $\sqrt{1-\gamma}$ in the superposition
\begin{equation}\label{eq:iqae-0}
    \mathcal{A}|0^n\rangle = \sqrt{1-\gamma} |0^n\rangle + \sqrt{\gamma} |\perp\rangle,
\end{equation}
where $\mathcal{A}$ is a composite unitary containing unitaries generating the NOQE ansatz states relevant to specific matrix elements between these, and $\langle0^n|\perp\rangle = 0$. The explicit form of $\mathcal{A}$ will be specified for both the overlap and Hamiltonian matrix element calculations below. An important difference between Eq. \eqref{eq:qae} and Eq. \eqref{eq:iqae-0} is that in Eq. \eqref{eq:iqae-0} both the initial state and the target state are $|0^n\rangle$. This means that their reflection operators are also both $\mathcal{S}_{\psi_0} = \mathcal{S}_{0} = I - 2\ket{0^n}\bra{0^n}$, which can be implemented with a multi-qubit controlled-$Z$ gate without any ancilla qubit. We therefore omit the additional ancilla qubit in Eq.~\ref{eq:iqae-0} and the subsequent analysis.

\subsection{Matrix elements in NOQE}

In NOQE, for each pair of non-orthogonal ansatz states $\ket{\psi_i}$ and $\ket{\psi_j}$,
we need to estimate the Hamiltonian and overlap matrix elements

\begin{equation}\label{eq:HS_def}
    H_{ij} = \langle \psi_i | H | \psi_j \rangle,
    \qquad
    S_{ij} = \langle \psi_i | \psi_j \rangle.
\end{equation}

The ansatz states are of the form $\ket{\psi_i}=e^{\hat{\tau}_i}\ket{\psi_{\mathrm{UHF}}}$,
where $\ket{\psi_{\mathrm{UHF}}}$ is the unrestricted Hartree--Fock (UHF) state.
Let $V$ be the unitary preparing $\ket{\psi_{\mathrm{UHF}}}$ from $\ket{0^n}$, i.e.,
$V\ket{0^n}=\ket{\psi_{\mathrm{UHF}}}$.
The key idea of IQAE--NOQE is to encode each matrix element as a measurement probability
(amplitude squared) of such a state-preparation unitary $\mathcal{A}$ acting on $\ket{0^n}$,
and then apply IQAE to estimate that probability with near-Heisenberg query scaling.

\subsection{Overlap matrix estimation via IQAE}\label{subsec:overlap_iqae}

We observe that
\begin{equation}\label{eq:S_amp_basic}
    V^{\dagger}e^{-\hat{\tau}_i}e^{\hat{\tau}_j}V\ket{0^n}
    = S_{ij}\ket{0^n}+\sqrt{1-|S_{ij}|^2}\ket{\perp},
\end{equation}
where $\ket{\perp}$ denotes a normalized state orthogonal to $\ket{0^n}$, i.e., $\langle 0^n \mid \perp \rangle = 0$. Because the relevant ansatz states lie in a fixed-particle-number sector orthogonal to $\ket{0^n}$, the desired matrix element is encoded directly in the $\ket{0^n}$ amplitude. As a result, unlike the standard IQAE construction, the present IQAE--NOQE formulation does not require an additional ancilla qubit for amplitude encoding. Running IQAE by treating $\mathcal{A} = V^{\dagger}e^{-\hat{\tau}_i}e^{\hat{\tau_j}}V$ can estimate $|S_{ij}|^2$ (and thus $|S_{ij}|$), leaving the phase of the complex $S_{ij}$ unresolved. In order to obtain both the real and imaginary parts of $S_{ij}$, we prepare four auxiliary states with $V_{\pm}$ and $V_{\pm i}$
\begin{equation}
\begin{aligned}
    V_\pm\ket{0^n} &= \frac{1}{\sqrt{2}}(\ket{0^n} \pm \ket{\psi_{\mathrm{UHF}}}) \\
    V_{\pm i}\ket{0^n} &= \frac{1}{\sqrt{2}}(\ket{0^n} \pm i \ket{\psi_{\mathrm{UHF}}}), 
\end{aligned}
\end{equation}
then using the fact that $e^{\hat{\tau}_i}$ preserves the particle number, which means $e^{\hat{\tau}_i}\ket{0^n} = \ket{0^n}$, we have the following
\begin{equation}\label{eq: S_QAE}
\begin{aligned}
    V_\pm^{\dagger}e^{-\hat{\tau}_i}e^{\hat{\tau}_j}V_\pm \ket{0^n} &= \alpha_\pm\ket{0^n} + \sqrt{1 - |\alpha_\pm|^2}\ket{\perp} \\
    V_-^{\dagger}e^{-\hat{\tau}_i}e^{\hat{\tau}_j}V_{\pm i} \ket{0^n} &= \alpha_{\pm i}\ket{0^n} + \sqrt{1 - |\alpha_{\pm i}|^2}\ket{\perp},
\end{aligned}
\end{equation}
where
\begin{equation}\label{eq: Re_Im_S}
\begin{aligned}
    |\alpha_\pm|^2 &= \frac{1}{4}\left(1+\left|S_{ij}\right|^2\pm2 \operatorname{Re}\left(S_{ij}\right)\right) \\
    |\alpha_{\pm i}|^2 &= \frac{1}{4}\left(1+\left|S_{ij}\right|^2\pm 2 \operatorname{Im}\left(S_{ij}\right)\right)
    .
\end{aligned}
\end{equation}
Running IQAE with $\mathcal{A}_{\pm}^{(S)} = V_\pm^{\dagger}e^{-\hat{\tau}_i}e^{\hat{\tau}_j}V_\pm$ and $\mathcal{A}_{\pm i}^{(S)} = V_{-}^{\dagger}e^{-\hat{\tau}_i}e^{\hat{\tau}_j}V_{\pm i}$ respectively as in Eq. \eqref{eq: S_QAE} would then estimate $|\alpha_\pm|^2$ and $|\alpha_{\pm i}|^2$, from which $\mathrm{Re}(S_{ij})$ and $\mathrm{Im}(S_{ij})$ can be solved using Eq. \eqref{eq: Re_Im_S}. This approach is summarized in Algorithm \ref{alg:EstimateOverlap}, where $\operatorname{IQAE}(\mathcal{A},\varepsilon,\delta)$ denotes running
Algorithm~\ref{alg:iqae} with state-preparation unitary $\mathcal{A}$, target
accuracy $\varepsilon$, and failure probability $\delta$.

\begin{algorithm}[H]
\caption{\textsc{EstimateOverlap}}
\label{alg:EstimateOverlap}
\begin{algorithmic}[1]
\Require
UCC ansatz and auxiliary unitaries; ansatz state indices $i,j$; IQAE parameters $(\varepsilon,\delta)$
\Ensure
Estimate $\hat{S}_{ij}$ of $S_{ij}$

\State Construct state preparation unitaries $\mathcal{A}_{\pm}^{(S)} = V_\pm^{\dagger}e^{-\hat{\tau}_i}e^{\hat{\tau}_j}V_\pm$ and $\mathcal{A}_{\pm i}^{(S)} = V_-^{\dagger}e^{-\hat{\tau}_i}e^{\hat{\tau}_j}V_{\pm i}$

\State $|\alpha_\pm|^2 \gets \mathrm{IQAE}(\mathcal{A}_\pm^{(S)}, \varepsilon, \delta)$, $|\alpha_{\pm i}|^2 \gets \mathrm{IQAE}(\mathcal{A}_{\pm i}^{(S)}, \varepsilon, \delta)$

\State Solve for $\mathrm{Re}(S_{ij})$ and $\mathrm{Im}(S_{ij})$ using Eq. \eqref{eq: Re_Im_S}
\State \Return $\widehat{S}_{ij} = \mathrm{Re}(S_{ij}) + i \mathrm{Im}(S_{ij})$
\end{algorithmic}
\end{algorithm}

\subsection{Hamiltonian matrix estimation via IQAE}\label{subsec:ham_iqae}

\begin{algorithm}[H]
\caption{\textsc{EstimatePauliExpect}}
\label{alg:EstimatePauli}
\begin{algorithmic}[1]
\Require
UCC ansatz and auxiliary unitaries, Pauli word $P_{\ell}$, ansatz state indices $i$ and $j$, IQAE parameters ($\varepsilon$, $\delta$)
\Ensure
Estimate $\hat{P}_{ij,\ell}$ of $P_{ij,\ell}=\bra{\psi_i}P_{\ell}\ket{\psi_j}$

\State Construct state preparation unitaries $\mathcal{A}_\pm^{(P_\ell)} = V_\pm^{\dagger}e^{-\hat{\tau}_i}P_\ell e^{\hat{\tau}_j}V_\pm$ and $\mathcal{A}_{\pm i}^{(P_\ell)} = V_-^{\dagger}e^{-\hat{\tau}_i}P_\ell e^{\hat{\tau}_j}V_{\pm i}$

\State $|\beta_\pm|^2 \gets \mathrm{IQAE}(\mathcal{A}_\pm^{(P_\ell)}, \varepsilon, \delta)$, $|\beta_{\pm i}|^2 \gets \mathrm{IQAE}(\mathcal{A}_{\pm i}^{(P_\ell)}, \varepsilon, \delta)$

\State Solve for $\mathrm{Re}(P_{ij,\ell})$ and $\mathrm{Im}(P_{ij,\ell})$ using Eq. \eqref{eq: Re_Im_H}
\State \Return $\widehat{P}_{ij} = \mathrm{Re}(P_{ij,\ell}) + i \mathrm{Im}(P_{ij,\ell})$
\end{algorithmic}
\end{algorithm}

A similar approach can be used for Hamiltonian matrix estimation.
Using the Jordan--Wigner transformation \cite{whitfield2011simulation,jw}, the electronic Hamiltonian can be decomposed into a weighted
sum of Pauli operators $H=\sum_{\ell=1}^{L} c_\ell P_\ell$.
We calculate each Pauli term $P_{ij,\ell}=\bra{\psi_i}P_\ell\ket{\psi_j}$ on quantum hardware and then assemble
\begin{equation}
    H_{ij}=\bra{\psi_i}H\ket{\psi_j}=\sum_{\ell=1}^{L} c_\ell\, P_{ij,\ell}.
\end{equation}

For conceptual clarity, the presentation below treats the Pauli terms individually. In practice, however, the expectation values of these $L$ Pauli terms need not be obtained from $L$ distinct measurement settings: more general measurement-reduction strategies can allow multiple Pauli expectations to be inferred from a smaller set of measured observables \cite{huggins2021efficient}. Such measurement-reduction strategies are compatible with the present framework and can be incorporated on top of the IQAE-based estimation procedure.

To proceed, we notice that
\begin{equation}\label{eq: H_QAE}
\begin{aligned}
    V_\pm^{\dagger}e^{-\hat{\tau}_i}P_\ell e^{\hat{\tau}_j}V_\pm \ket{0^n}
    &= \beta_\pm\ket{0^n} + \sqrt{1 - |\beta_\pm|^2}\ket{\perp}, \\
    V_{-}^{\dagger}e^{-\hat{\tau}_i} P_\ell e^{\hat{\tau}_j}V_{\pm i} \ket{0^n}
    &= \beta_{\pm i}\ket{0^n} + \sqrt{1 - |\beta_{\pm i}|^2}\ket{\perp},
\end{aligned}
\end{equation}
where
\begin{equation}\label{eq: Re_Im_H}
\begin{aligned}
    |\beta_\pm|^2 &= \frac{1}{4}\left|\langle 0^n| P_\ell|0^n\rangle\pm\langle 0^n| P_\ell\left|\psi_j\right\rangle\pm\left\langle\psi_i\right| P_\ell|0^n\rangle+P_{ij,\ell} \right|^2, \\
    |\beta_{\pm i}|^2 &= \frac{1}{4}\left|\langle 0^n| P_\ell|0^n\rangle\pm i\langle 0^n| P_\ell\left|\psi_j\right\rangle-\left\langle\psi_i\right| P_\ell|0^n\rangle\mp i P_{ij,\ell} \right|^2.
\end{aligned}
\end{equation}

Here we used the asymmetry of $V_-$ and $V_{\pm i}$ to learn the phases of $P_{ij,\ell}$ in Eq. \eqref{eq: H_QAE}. Running IQAE with state-preparation operators
$\mathcal{A}_\pm^{(P_\ell)} = V_\pm^{\dagger}e^{-\hat{\tau}_i}P_\ell e^{\hat{\tau}_j}V_\pm$
and
$\mathcal{A}_{\pm i}^{(P_\ell)} = V_-^{\dagger}e^{-\hat{\tau}_i}P_\ell e^{\hat{\tau}_j}V_{\pm i}$
allows estimation of
$|\beta_\pm|^2$ and $|\beta_{\pm i}|^2$. Each application of the unitary $\mathcal{A}$ thus requires two circuit primitives taken from the set of ansatz state preparation operators and their adjoints, $\{e^{\pm\hat{\tau}_i}\}$.
We also note that $\langle0^n|P_\ell\ket{0^n}$ can be computed efficiently classically, and
$\langle0^n|P_\ell\ket{\psi_j}=\langle0^n|P_\ell e^{\hat{\tau}_j}V\ket{0^n}$ can be learned by applying IQAE
to $\mathcal{A}_\pm = P_\ell e^{\hat{\tau}_j}V_\pm$ and $\mathcal{A}_{\pm i} = P_\ell e^{\hat{\tau}_j}V_{\pm i}$,
following the same procedure as in the overlap estimation in Section~\ref{subsec:overlap_iqae} above.

With these quantities available, Eq. \eqref{eq: Re_Im_H} then provides sufficient information to solve for both the real and imaginary parts of the Hamiltonian matrix Pauli components $P_{ij,\ell}$. This procedure for estimating $P_{ij,\ell}$ is summarized in Algorithm \ref{alg:EstimatePauli}.

\begin{algorithm}[H]
\caption{IQAE--NOQE: Estimating overlap and Hamiltonian matrices}
\label{alg:iqae-noqe-SH}
\begin{algorithmic}[1]
\Require
UCC ansatz and auxiliary unitaries,
Hamiltonian decomposition $H=\sum_{\ell=1}^{L} c_\ell P_\ell$ (Pauli words $P_\ell$),
IQAE parameters $(\varepsilon,\delta)$
\Ensure
Estimated overlap matrix $\widehat{\mathbf{S}}$ and Hamiltonian matrix $\widehat{\mathbf{H}}$
\For{$1 \le i \le j \le M$}
    \State $\widehat{S}_{ij} \gets \textsc{EstimateOverlap}(i,j,\varepsilon, \delta)$
    \For{$\ell = 1$ to $L$}
        \State $\widehat{P}_{ij,\ell} \gets \textsc{EstimatePauliExpect}(P_\ell, i, j, \varepsilon, \delta)$
        \State $\widehat{H}_{ij} \gets \widehat{H}_{ij} + c_\ell\, \widehat{P}_{ij,\ell}$
    \EndFor
\EndFor
\State \Return $\widehat{\mathbf{S}},\,\widehat{\mathbf{H}}$
\Comment{solve $\widehat{\mathbf{H}}\mathbf{x}=E\,\widehat{\mathbf{S}}\mathbf{x}$ to get energies}
\end{algorithmic}
\end{algorithm}

\subsection{Full IQAE--NOQE pipeline and resources}

\begin{figure*}[!t]
    \centering
    \includegraphics[width=0.7\linewidth]{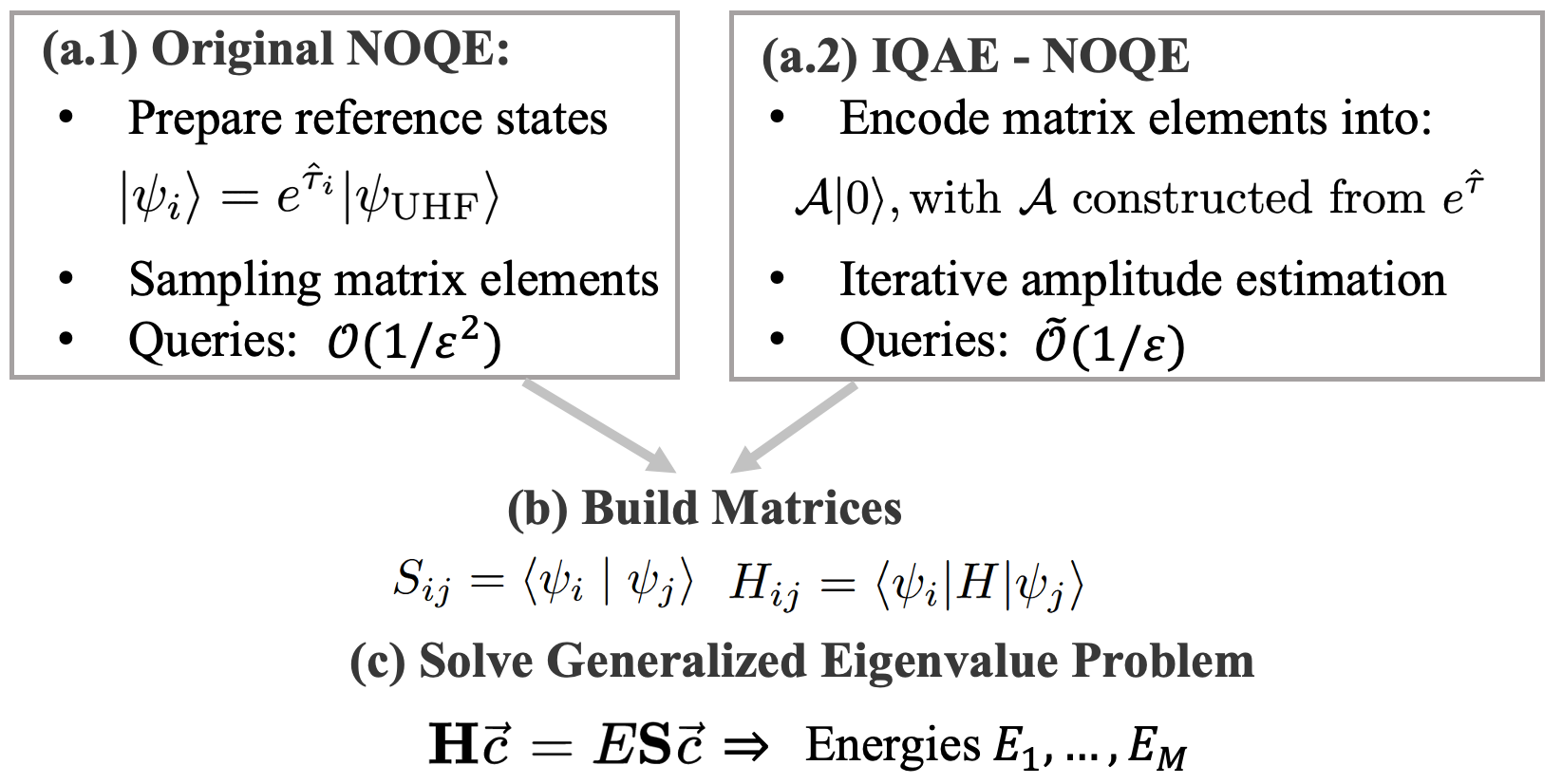}
    \caption{Overview of the NOQE and IQAE--NOQE workflows.
    (a.1) In the original NOQE protocol, non-orthogonal ansatz states
    $\ket{\psi_i}=e^{\hat{\tau}_i}\ket{\psi_{\mathrm{UHF}}}$ are prepared and the
    overlap and Hamiltonian matrix elements between these states are estimated via statistical sampling,
    leading to a measurement cost scaling as $\mathcal{O}(1/\varepsilon^2)$.
    (a.2) In IQAE--NOQE, the same matrix elements are coherently encoded as amplitudes
    of a quantum state $\mathcal{A}\ket{0^n}$, with $\mathcal{A}$ constructed from
    state-preparation unitaries, and are estimated using iterative quantum amplitude
    estimation, achieving a near-Heisenberg scaling of $\tilde{\mathcal{O}}(1/\varepsilon)$.
    (b) The estimated overlap and Hamiltonian matrices are assembled.
    (c) Solving the resulting generalized eigenvalue problem on a classical computer yields the energies. 
    }

    \label{fig:diagram}
\end{figure*}

\begin{figure}[!t]
    \centering
    \includegraphics[width=\linewidth]{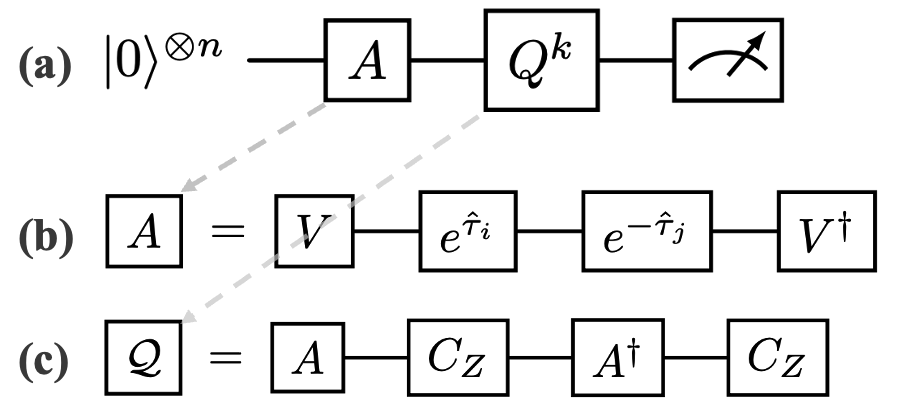}
    \caption{Circuit structure for IQAE--NOQE.
    (a) Standard IQAE setting, where an
    operator $\mathcal{A}$ prepares a state encoding the target amplitude, followed
    by repeated applications of the Grover operator $\mathcal{Q}^k$ and measurement.
    (b) Construction of the state-preparation operator $\mathcal{A}$ for NOQE
    matrix-element estimation.
    Here $\mathcal{A}$ is built from the NOQE ansatz-state preparation and
    auxiliary unitaries, typically of the form
    $V\,e^{\hat{\tau}_i}\,e^{-\hat{\tau}_j}\,V^\dagger$, such that the desired overlap
    or Hamiltonian matrix element is encoded into the measurement probability.
    (c) Definition of the Grover operator
    $\mathcal{Q} = \mathcal{A}\,C_Z\,\mathcal{A}^\dagger\,C_Z$,
    which amplifies the encoded amplitude and enables near-Heisenberg scaling in
    the estimation of matrix elements. Here $C_Z$ denotes a multi-qubit controlled-$Z$ operation. 
    Each application of the Grover operator $\mathcal{Q}$ thus requires two applications of $\mathcal{A}$
 and hence four circuit primitives
taken from the set of ansatz state preparation operators and their adjoints, $\{ e^{ \pm \hat{\tau}_i } \}$. }
    \label{fig:circuit}
\end{figure}

After estimating $S_{ij}$ and $H_{ij}$ for each pair of ansatz states $\ket{\psi_i}$ and $\ket{\psi_j}$
and then assembling $\mathbf{S}$ and $\mathbf{H}$, we solve the generalized eigenvalue problem
$\mathbf{H}\vec{c} = \mathbf{E}\mathbf{S}\vec{c}$ on a classical processor to obtain electronic energies. The complete procedure is summarized in Algorithm~\ref{alg:iqae-noqe-SH}, which we refer to as IQAE--NOQE.
The pipeline of IQAE--NOQE is also schematically shown in Fig.~\ref{fig:diagram} and a high-level circuit is
provided in Fig.~\ref{fig:circuit}.

In the implementation of IQAE--NOQE, the initial state and the target state
(whose amplitude encodes $S_{ij}$ or $P_{ij,\ell}$) are both $\ket{0^n}$.
Therefore, in the construction of the Grover operator $\mathcal{Q}$, the reflection operators are simply
$\mathcal{S}_{\psi} = \mathcal{S}_{\psi_0} = I - 2 \ket{0^n}\bra{0^n}$.
This is equivalent to a multi-qubit controlled-$Z$ gate and can be implemented efficiently using
$\mathcal{O}(n)$ two-qubit gates~\cite{zindorf2024efficient}. In addition, for each pair of ansatz states, the estimation of the overlap matrix requires 4 circuits ($\mathcal{A}^{(S)}_\pm$ and $\mathcal{A}^{(S)}_{\pm i}$), and the estimation of each Pauli term requires 12 circuits ($\mathcal{A}^{(P_\ell)}_\pm$, $\mathcal{A}^{(P_\ell)}_{\pm i})$, and 4 circuits for each  $\langle0^n|P_\ell \ket{\psi_i}$ and $\langle0^n|P_\ell \ket{\psi_j}$), meaning that the circuit overhead is constant, i.e., independent of the qubit number $N$.

We note that the additive error $\varepsilon$ in the estimated amplitudes is dimensionless and translates into an additive error of $\mathcal{O}(\varepsilon)$ in the estimated overlap matrix elements and individual Pauli matrix elements, which are then combined to reconstruct the Hamiltonian matrix elements.

We emphasize that the $\tilde{\mathcal{O}}(1/\varepsilon)$ scaling provided by IQAE
arises from the statistical structure of amplitude estimation and is
independent of the molecular system size $N$ or the number of ansatz states.
In particular, while the overall resource requirements still scale with the
number of matrix elements to be estimated, the quadratic improvement in
precision with respect to $\varepsilon$ holds uniformly across system sizes.
Consequently, the measurement-efficiency advantage of IQAE--NOQE persists for
larger systems and expanded ansatz spaces.

Overall, IQAE--NOQE retains a circuit structure comparable to sampling-based NOQE, but replaces incoherent sampling with coherent Grover amplification.
As a result, IQAE--NOQE achieves near-Heisenberg query scaling $\tilde{\mathcal{O}}(1/\varepsilon)$ in the number of
queries to the most expensive subroutine (dominated here by the UCC-type circuit blocks that generate the dynamic correlations of the ansatz, $e^{\hat{\tau}_i}$),
at the cost of moderately increased circuit depth due to the repeated applications of $\mathcal{Q}$.

\subsection{Intuitive origin of the IQAE advantage for NOQE}\label{subsec:intuition_iqae_adv}

Here we give a brief intuitive explanation of the quadratic improvement offered by IQAE. It is easiest to understand this by comparing how information is accumulated in the two algorithms.

\paragraph{Sampling-based NOQE (incoherent accumulation of information).} In the original NOQE protocol, each matrix element is estimated from repeated projective measurements
(e.g., Hadamard-test style circuits and Pauli-term sampling). Each shot produces a single binary outcome.
Crucially, after measurement the quantum state is destroyed, so different repetitions cannot ``add their
information'' coherently. The estimate therefore improves only by averaging many independent samples.
By the law of large numbers, achieving additive error $\varepsilon$ requires $\mathcal{O}(1/\varepsilon^2)$
shots (and hence $\mathcal{O}(1/\varepsilon^2)$ queries to the relevant state-preparation blocks).

\paragraph{IQAE--NOQE (coherent accumulation of information before measurement).}
In contrast, the IQAE-based formulation accumulates information coherently before measurement.
By encoding the target quantity as a phase $\theta$, repeated applications of the Grover operator $\mathcal{Q}$ amplify this phase linearly with the number of iterations.
As a result, a fixed uncertainty in the final readout translates into an uncertainty that decreases as $\mathcal{O}(1/k)$ with the maximum Grover depth $k$.
The quadratic improvement thus stems from coherent amplitude accumulation prior to measurement, rather than from reducing projection noise in individual shots.

\paragraph{What is the real tradeoff?}

\begin{figure*}[t]
\centering

\begin{tabular}{c c c c c}
\toprule
Method & Statistical mechanism & Circuit depth & Total queries & Query complexity \\
\midrule
Original NOQE
& Sampling/averaging
& $D=\mathcal O(1)$
& $Q_{\rm NOQE}=1\times R,\quad R=\mathcal O(1/\varepsilon^2)$
& $\mathcal O(1/\varepsilon^2)$ \\

IQAE--NOQE
& Amplitude estimation
& $D_1,D_2,\ldots,D_{\max}=\mathcal O(1/\varepsilon)$
& $Q_{\rm IQAE}=\sum_i D_iR_i$
& $\widetilde{\mathcal O}(1/\varepsilon)$ \\
\bottomrule
\end{tabular}

\vspace{0.6cm}

\includegraphics[width=0.48\textwidth]{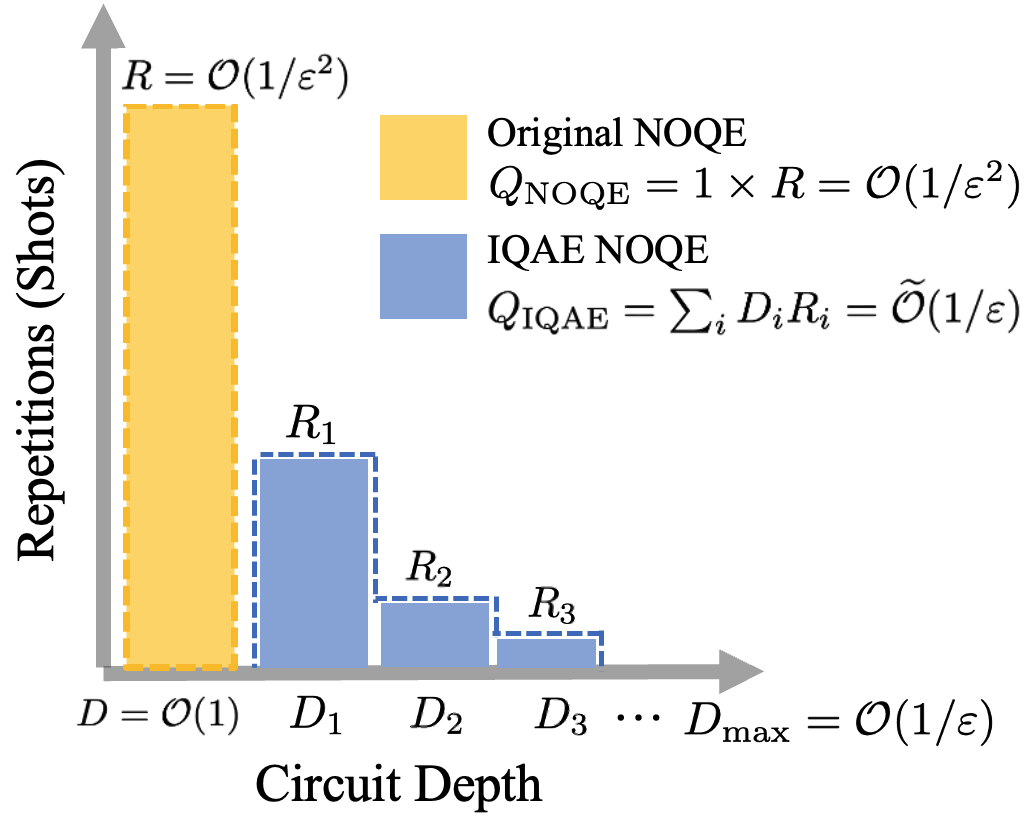}

\caption{
Conceptual resource comparison between the original sampling-based NOQE and
IQAE--NOQE. In the original NOQE protocol, each circuit execution uses the ansatz state-preparation blocks $e^{ \hat{\tau}_{i(j)}}$ for the two ansatz states $i(j)$ once, including their correlation operators and basis-rotation blocks. These can
be viewed as having a reference circuit depth $D=\mathcal{O}(1)$. The matrix elements
are then estimated by repeated sampling. Reaching additive precision
$\varepsilon$ requires $R=\mathcal{O}(1/\varepsilon^2)$ repetitions, giving
$Q_{\rm NOQE}=1\times R=\mathcal{O}(1/\varepsilon^2)$ total queries.
In IQAE--NOQE, the estimation is performed using a sequence of circuits with
different Grover depths $D_1,D_2,\ldots,D_{\max}$, where
$D_{\max}=\mathcal{O}(1/\varepsilon)$. If the circuit at depth $D_i$ is repeated
$R_i$ times, the total query count is
$Q_{\rm IQAE}=\sum_i D_iR_i=\widetilde{\mathcal{O}}(1/\varepsilon)$.
Thus IQAE--NOQE replaces many independent shallow repetitions with a smaller
set of coherently amplified circuits, reducing the total query complexity from
$\mathcal{O}(1/\varepsilon^2)$ to
$\widetilde{\mathcal{O}}(1/\varepsilon)$.
}

\label{fig:noqe_iqae_resource_comparison}
\end{figure*}

The advantage of IQAE is not that individual measurements become less noisy; rather, IQAE changes the workflow
from ``measure many times and average'' to ``amplify coherently, then measure.''
In the sampling-based setting, achieving additive error $\varepsilon$ requires $\mathcal{O}(1/\varepsilon^2)$ circuit repetitions, each involving a single query to the state-preparation routine, resulting in $\mathcal{O}(1/\varepsilon^2)$ total queries. In contrast, IQAE achieves near-Heisenberg query complexity $\tilde{\mathcal{O}}(1/\varepsilon)$ by coherently amplifying the signal before measurement.

This resource comparison is summarized schematically in Fig.~\ref{fig:noqe_iqae_resource_comparison}. IQAE--NOQE substantially reduces both the number of circuit repetitions and the total number of queries, at the cost of increased circuit depth due to repeated Grover iterations. Importantly, since the total query count effectively captures the overall computational effort (i.e., the ``computational area'' given by circuit depth times the number of executions), this represents not merely a redistribution of resources, but a genuine reduction in total computational cost when the query complexity is the dominant consideration, thereby achieving an overall improvement in resource efficiency.

With the IQAE--NOQE workflow established, we next validate the protocol on a concrete quantum-chemistry benchmark by applying it to the hydrogen molecule.

\section{Application to the Electronic States of the Hydrogen Molecule}\label{sec:result}

\begin{figure}[htbp]
    \centering
    \includegraphics[width=\linewidth]{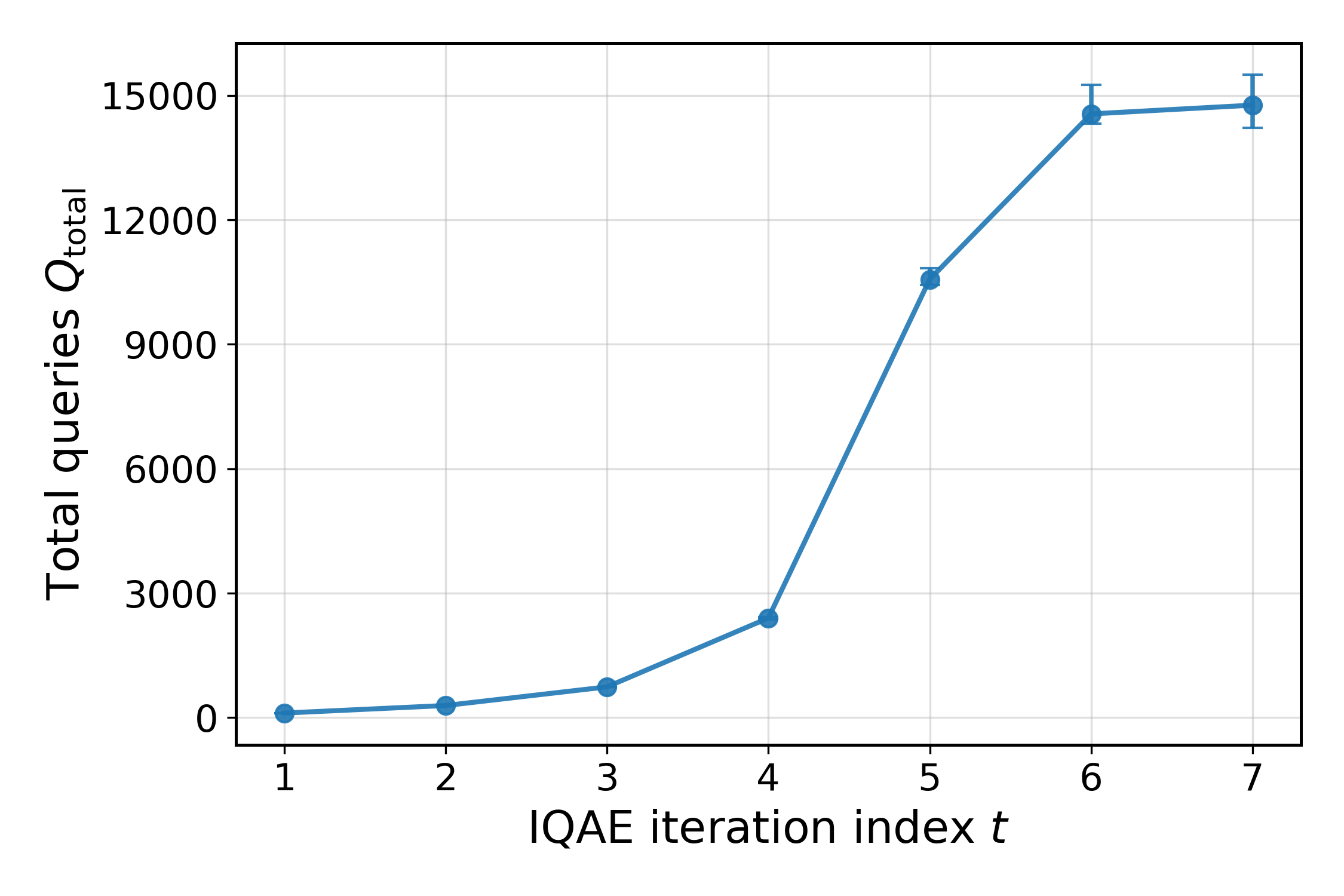}
    \caption{Cumulative total query count $Q_{\mathrm{total}}$ as a function of the IQAE iteration index $t$ for the full H$_2$ numerical experiment. Here $Q_{\mathrm{total}}(t)$ denotes the accumulated number of IQAE queries up to iteration $t$, summed over all IQAE subroutines entering the overlap and Hamiltonian matrix reconstruction, rather than the query count for a single matrix element. At each iteration index $t$, we report the median cumulative query count across the trials that reached iteration $t$, and the error bars denote the interquartile range (25\%--75\%). The plot ends at $t=7$ because no trial in this set of experiments required more than seven IQAE iterations, demonstrating a stable and reproducible stopping behavior of the adaptive IQAE procedure. The stepwise increase reflects the iterative refinement structure of IQAE.}
    \label{fig:queries vs iteration}
\end{figure}

\begin{figure*}[htbp]
    \centering
    \begin{subfigure}[b]{0.32\textwidth}
        \centering
        \includegraphics[width=\linewidth]{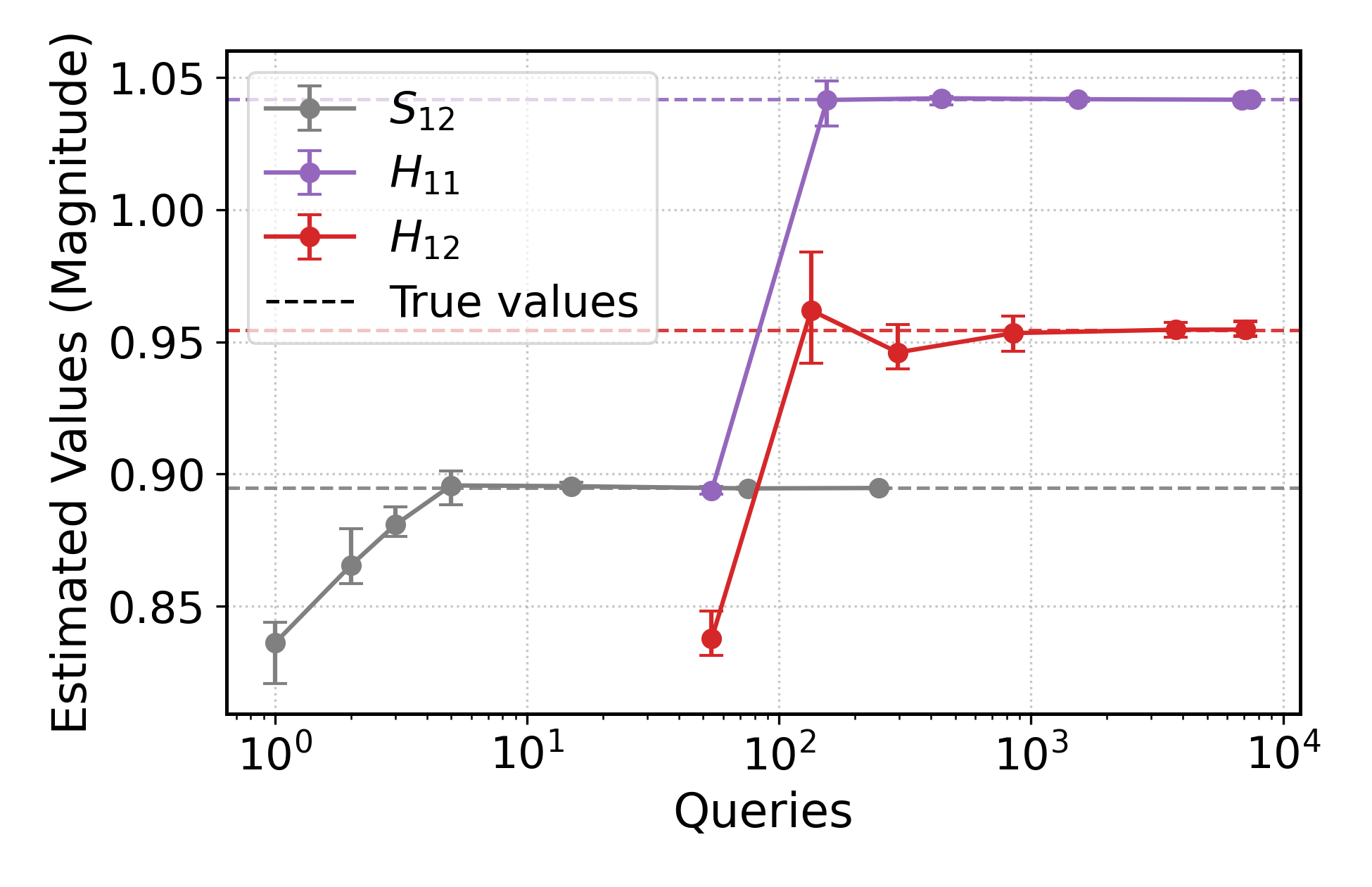}
        \caption{Accurate matrix element estimation}
        \label{fig:three_panel1}
    \end{subfigure}
    \hfill
    \begin{subfigure}[b]{0.32\textwidth}
        \centering
        \includegraphics[width=\linewidth]{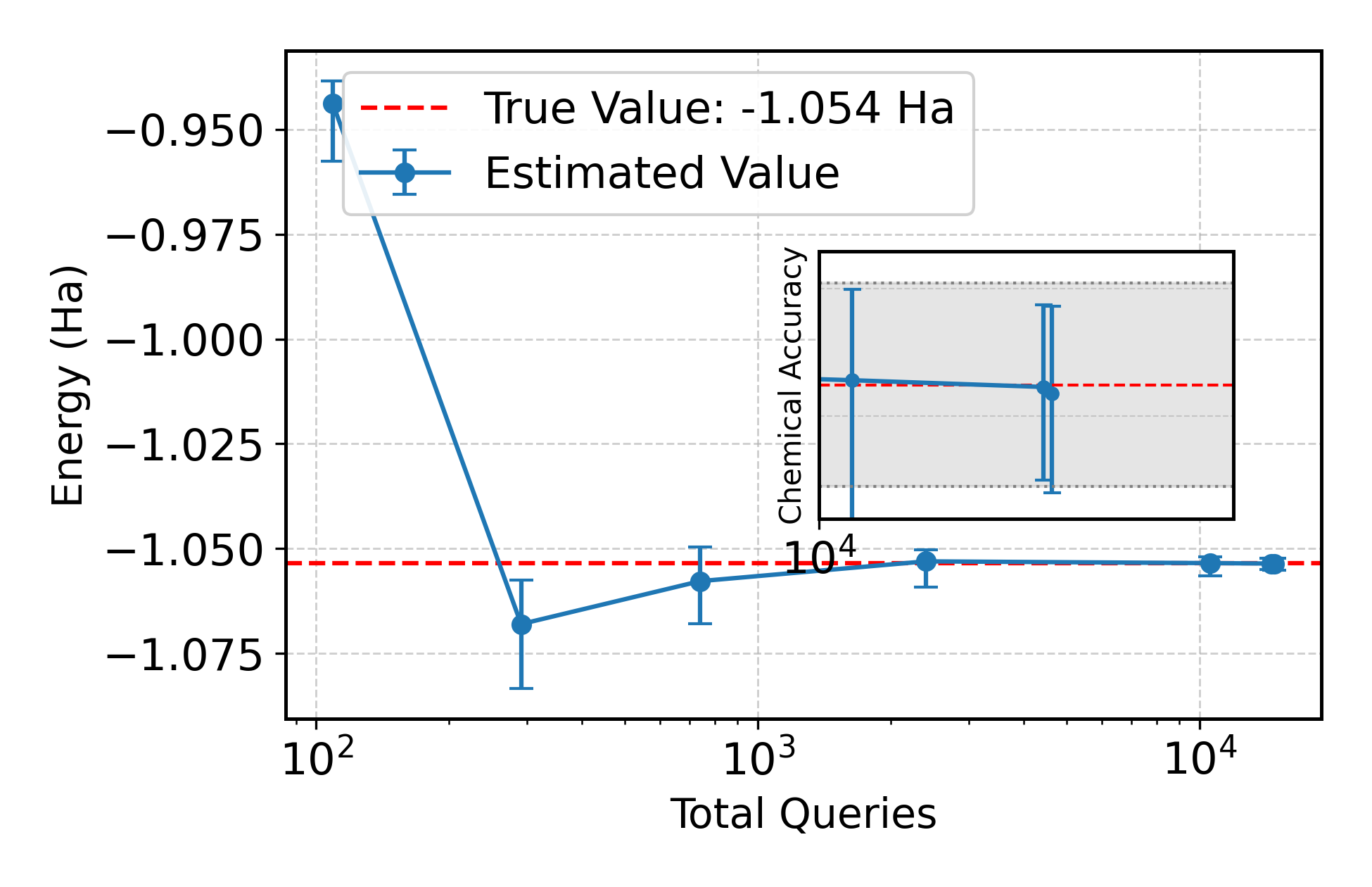}
        \caption{Energy convergence}
        \label{fig:three_panel2}
    \end{subfigure}
    \hfill
    \begin{subfigure}[b]{0.32\textwidth}
        \centering
        \includegraphics[width=\linewidth]{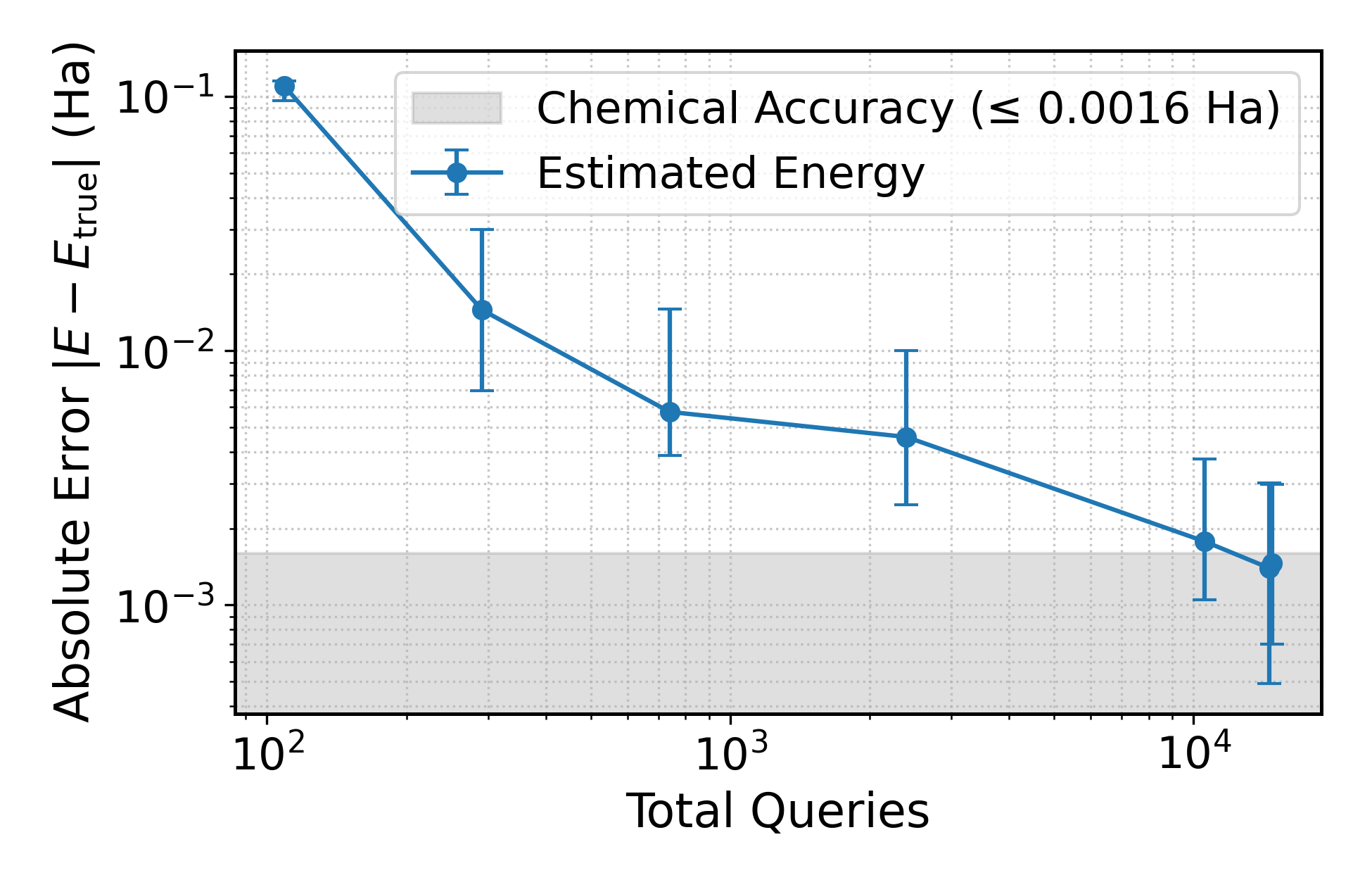}
        \caption{Chemical accuracy achievement}
        \label{fig:three_panel3}
    \end{subfigure}
    \caption{
    Performance of IQAE--NOQE for the hydrogen molecule as a function of the total number of  queries.
    Error bars indicate the interquartile range ($25\%$--$75\%$) over repeated trials, and dashed lines denote true values.
    (a) Convergence of estimated matrix elements, including the overlap $s_{12}$ and Hamiltonian elements $h_{11}$ and $h_{12}$.
    (b) Convergence of the estimated ground-state energy $E$; the inset shows a zoomed-in view of the high-query regime, where the energy estimate stabilizes within chemical accuracy relative to the true value.
    (c) Absolute energy error $|E - E_{\mathrm{true}}|$ versus total queries on a logarithmic scale.
    The shaded region denotes chemical accuracy ($\le 0.0016$~Ha).
    The overall reduction of the error with the total number of queries indicates increasingly accurate energy estimation enabled by iterative amplitude estimation. Overall, these results validate the correctness of the IQAE--NOQE algorithm.}
    \label{fig:three_panel}
\end{figure*}

We validate the accuracy and efficiency of our protocol by applying it to the hydrogen molecule ($\mathrm{H}_2$) at the Coulson-Fischer point (interatomic spacing = 1.2~\r{A}), a regime characterized by electronic spin-symmetry breaking and strong correlation \cite{coulson1949xxxiv}. This geometry serves as a stringent benchmark for the protocol, so successful energy estimation in this regime provides strong evidence for the robustness and query efficiency of IQAE--NOQE.

We use the STO-3G basis \cite{hehre1969self} for a minimal orbital representation, leading to a UHF reference state $|\psi_{\mathrm{UHF}}\rangle = |1100\rangle$ and two ansatz states $|\psi_1\rangle = e^{\hat{\tau}_1} |\psi_{\mathrm{UHF}}\rangle$ and $|\psi_2\rangle = e^{\hat{\tau}_2}|\psi_{\mathrm{UHF}}\rangle$. As noted earlier, the cluster operators $e^{\hat{\tau}_1}$ and $e^{\hat{\tau}_2}$
include both the UCCD excitation operators and the basis rotations needed to express
the two ansatz states in a common basis. Each state encodes four spin orbitals into four qubits. Under these conditions, the Hamiltonian and overlap matrices have the form
\begin{equation}\label{eq: hydrogen_matrices}
\mathbf{H}=\begin{pmatrix}
h_{11} & h_{12} \\
h_{12}^* & h_{22}
\end{pmatrix}, \quad
\mathbf{S}=\begin{pmatrix}
1 & s_{12} \\
s_{12}^* & 1
\end{pmatrix}.
\end{equation}

Here $h_{22}=h_{11}$ in the present symmetric H$_2$ setup, so the values to be estimated are $s_{12}$, $h_{12}$, and $h_{11}$. Solving the generalized eigenvalue problem in Eq.~\eqref{eq:gep} provides both the ground and excited state energies. Here, we focus on the ground state energy, a primary target in quantum chemistry. We compare against the noiseless simulation results of the original NOQE protocol as a ground-truth reference (see Appendix and Refs.~\cite{baek2023say,ren2026error} for NOQE implementation details).

In the IQAE--NOQE simulations, we set the target precision of estimating the amplitudes to
$\varepsilon = 10^{-3}$ and the confidence parameter to $\delta = 0.01$,
corresponding to an estimation error below $10^{-3}$, with failure
probability less than $1\%$. The quantities estimated by IQAE include
the overlap matrix element $s_{12}$ and Pauli strings associated with Hamiltonian matrix elements, which are subsequently
assembled to obtain the full Hamiltonian matrix elements.

In the present implementation, the four-qubit STO-3G Hamiltonian obtained after the Jordan--Wigner transformation contains 27 ungrouped Pauli terms. The overlap matrix element $s_{12}$ is estimated through a dedicated IQAE subroutine, while the Hamiltonian matrix elements are reconstructed from term-by-term IQAE estimates of the Pauli contributions entering the diagonal and off-diagonal matrix elements.

We emphasize that the present numerical study is intended as a proof-of-principle demonstration of IQAE-based matrix-element reconstruction, rather than a fully optimized accounting of NOQE measurement cost. In particular, we do not incorporate Pauli-term grouping or an optimized shot-allocation strategy across measurement settings. Instead, the Pauli contributions are estimated term by term, and each IQAE subroutine is run until convergence for the corresponding observable. The total query counts reported below should therefore be interpreted as implementation-level query counts for this explicit construction.

At a global iteration index $t$, the energy is reconstructed from the most recent available estimates of $s_{12}$, $h_{11}$, and $h_{12}$, after which the generalized eigenvalue problem is solved to obtain the ground-state energy. The cumulative total query count reported in the figures is defined as
\[
Q_{\mathrm{total}}(t)=Q_{s_{12}}(t)+Q_{h_{11}}(t)+Q_{h_{12}}(t),
\]
where each term denotes the cumulative IQAE query count of the corresponding subproblem up to iteration $t$. In our implementation, the cumulative query count of each IQAE subroutine is recorded as
\begin{equation}
    Q(t)=\sum_{r\le t}n_r(2k_r+1),
\end{equation}
where $k_r$ is the Grover power used at the $r$th IQAE update and $n_r$ is the number of repetitions performed at that update.

To characterize both the accuracy and resource usage of the algorithm, we perform 100 independent executions of the complete IQAE--NOQE protocol. Due to the adaptive stopping nature of IQAE, different trials may terminate after different numbers of Grover iterations. To enable a meaningful comparison across trials while preserving the intrinsic stepwise structure of the IQAE procedure, we adopt an iteration-index-based aggregation strategy.

Specifically, for each IQAE iteration index $t$, we collect all records corresponding to that iteration across independent trials and treat the accumulated number of queries up to that iteration, $Q_{\mathrm{total}}(t)$, as a trial-dependent quantity. We summarize the typical query cost at iteration $t$ by reporting the median over trials, while the interquartile range (25\%--75\%) captures the variability arising from stopping and measurement randomness. The estimated matrix elements and estimated energies at the same iteration index are aggregated in the same way.

With this aggregation scheme, the reported median $Q_{\mathrm{total}}(t)$ should be interpreted as the typical cumulative number of queries required to reach iteration $t$, rather than the query cost of any specific trial. This approach provides a robust characterization of both estimator convergence and resource growth.

Before presenting the results, we note that throughout this section, the ``true'' values of matrix elements and
energies are taken from noiseless simulations of the original NOQE
protocol. All circuit-level simulations of IQAE--NOQE are performed
using Qiskit \cite{Qiskit2024} and include shot noise from finite sampling, but do
not include gate or decoherence noise, allowing us to isolate the
statistical performance of the measurement protocol to better study the measurement issue.

\textbf{Performance of IQAE-NOQE for H$_2$.}\\
We first characterize the convergence behavior of the algorithm in 
Fig.~\ref{fig:queries vs iteration}. This shows the growth of the total number of
queries as a function of the number of IQAE iterations. We observe that all trials converge
within seven iterations, indicating a stable and reproducible stopping behavior
of the adaptive IQAE procedure. The early iterations produce only coarse
estimates and are therefore obtained at low query cost, while the majority of
the total cost is incurred as the algorithm approaches the target precision in later iterations.
This behavior is consistent with the designed operation of IQAE and confirms
that the protocol follows the expected adaptive refinement process.

Figure~\ref{fig:three_panel} summarizes the performance of IQAE--NOQE as
a function of the total number of queries. In
Fig.~\ref{fig:three_panel1}, we show the convergence of the estimated
matrix elements, including the overlap $s_{12}$ and Hamiltonian elements
$h_{11}$ and $h_{12}$. All estimated quantities approach their true
values within the target precision as the number of queries increases,
demonstrating the correctness of the matrix-element estimation step.

Figure~\ref{fig:three_panel2} shows the corresponding convergence of the
ground-state energy obtained by solving the generalized eigenvalue
problem using the estimated matrices. As the matrix elements converge,
the energy estimate stabilizes and enters the chemical-accuracy regime, as highlighted in the inset, which zooms in on the high-query region ($Q_{\mathrm{total}}>10^{4}$).

Finally, Fig.~\ref{fig:three_panel3} presents the absolute energy error
$|E - E_{\mathrm{true}}|$ as a function of the total number of queries on a logarithmic scale. The overall reduction of the energy
error provides direct evidence that the
iterative amplitude estimation procedure systematically improves the
energy estimate. Together, these results validate both the correctness
and the numerical stability of the IQAE--NOQE protocol.

\textbf{Comparison with original NOQE measurement protocol.}

\begin{figure*}[htbp]
    \centering
    \begin{subfigure}[b]{0.48\textwidth}
        \centering
        \includegraphics[width=\textwidth]{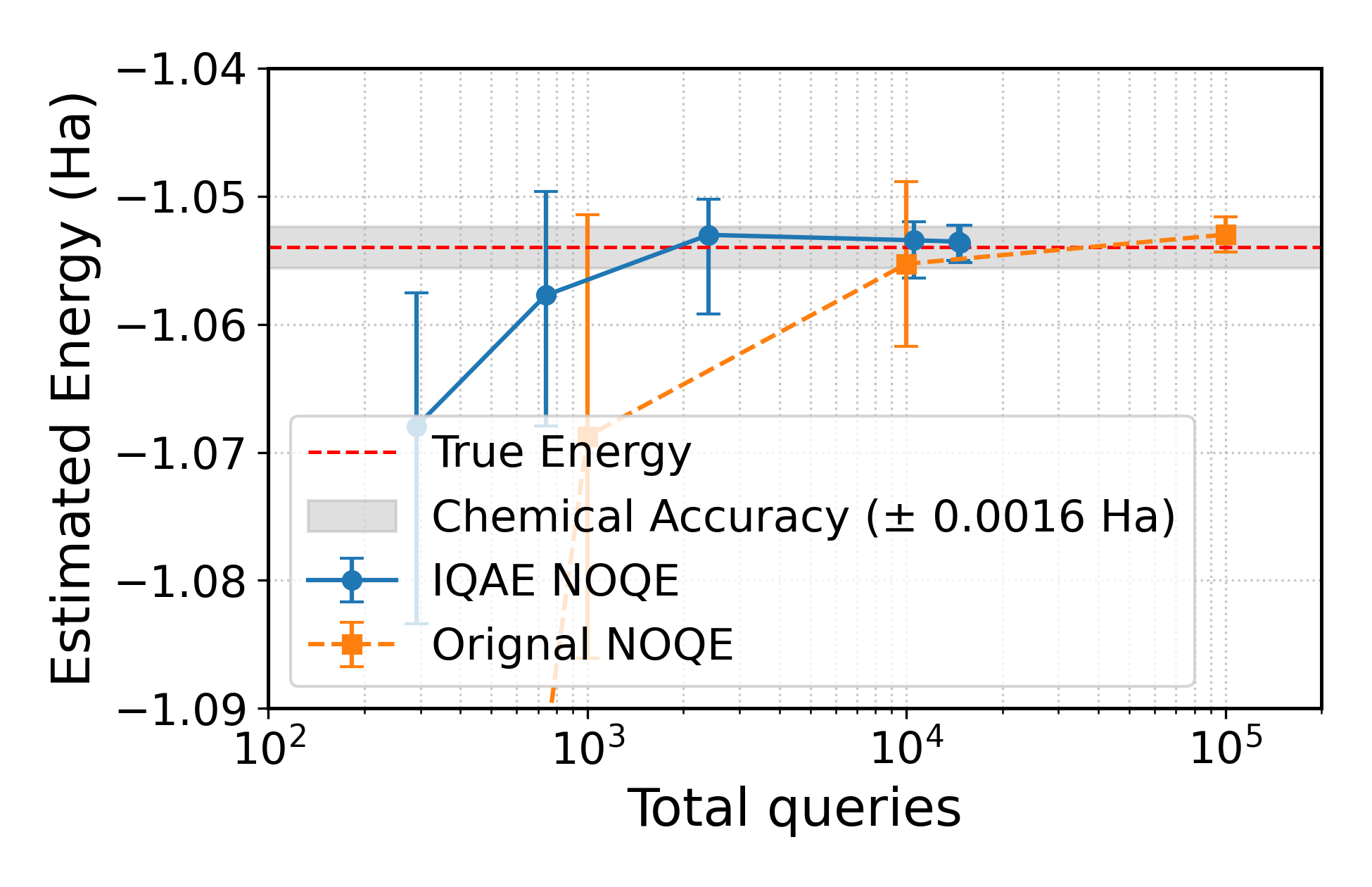}
        \caption{Energy convergence comparison}
        \label{fig:a}
    \end{subfigure}
    \hfill
    \begin{subfigure}[b]{0.48\textwidth}
        \centering
        \includegraphics[width=\textwidth]{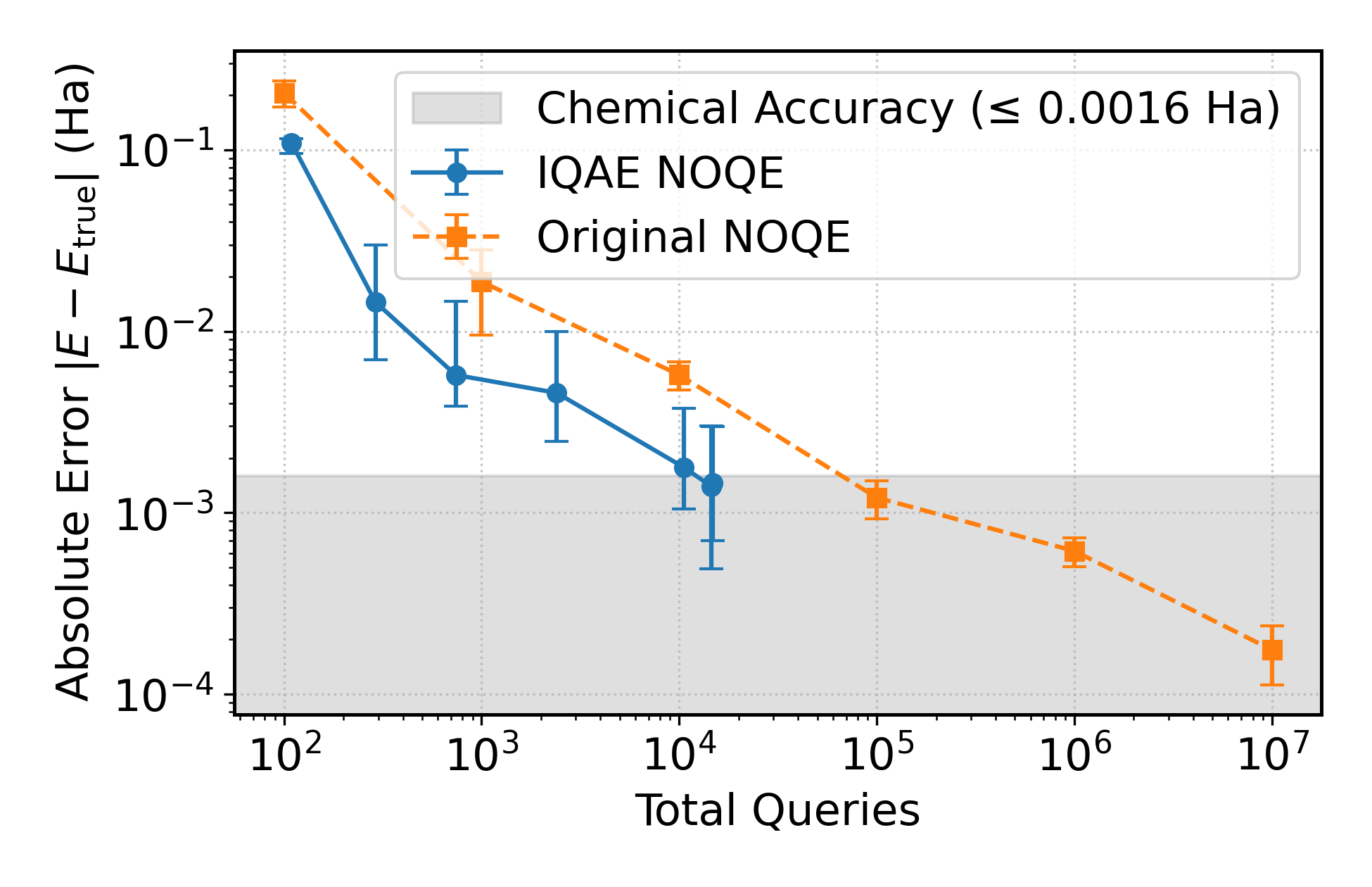}
        \caption{Energy error comparison}
        \label{fig:b}
    \end{subfigure}
    \caption{
    Comparison of energy estimation performance between IQAE--NOQE and the original NOQE.
    (a) IQAE--NOQE converges to the true energy with fewer queries and
    smaller statistical uncertainty than the original NOQE, indicating more stable and reliable energy estimation.
    (b) IQAE--NOQE reaches chemical accuracy with fewer queries than
    the original NOQE, demonstrating the improved query efficiency enabled by iterative amplitude estimation.
    }
    \label{fig:both}
\end{figure*}

We now compare the performance of IQAE--NOQE with that of the original
sampling-based measurement protocol of NOQE in Ref.~\cite{baek2023say}, focusing on the convergence behavior and the corresponding estimation error. Both protocols use the same ansatz states and electronic Hamiltonian for H$_2$ described above.

The results are summarized in Fig.~\ref{fig:both}. Panel (a)
plots the estimated ground-state energy as a function of the
total number of queries. IQAE--NOQE approaches the true energy value more rapidly
than the original NOQE and exhibits smaller statistical fluctuations at
comparable query counts. This indicates that IQAE--NOQE leads to a more stable and reliable convergence behavior.

Panel (b) of Fig.~\ref{fig:both} shows the absolute energy error $|E - E_{\mathrm{true}}|$
versus the total number of queries. IQAE--NOQE reaches chemical accuracy with
substantially fewer queries than the original NOQE, about one order of magnitude less, directly demonstrating the
improved query efficiency enabled by iterative amplitude estimation. Together, these results confirm that IQAE--NOQE provides a clear advantage in
measurement efficiency while preserving the accuracy of the original NOQE
framework.

\begin{figure}[htbp]
    \centering
    \includegraphics[width=\linewidth]{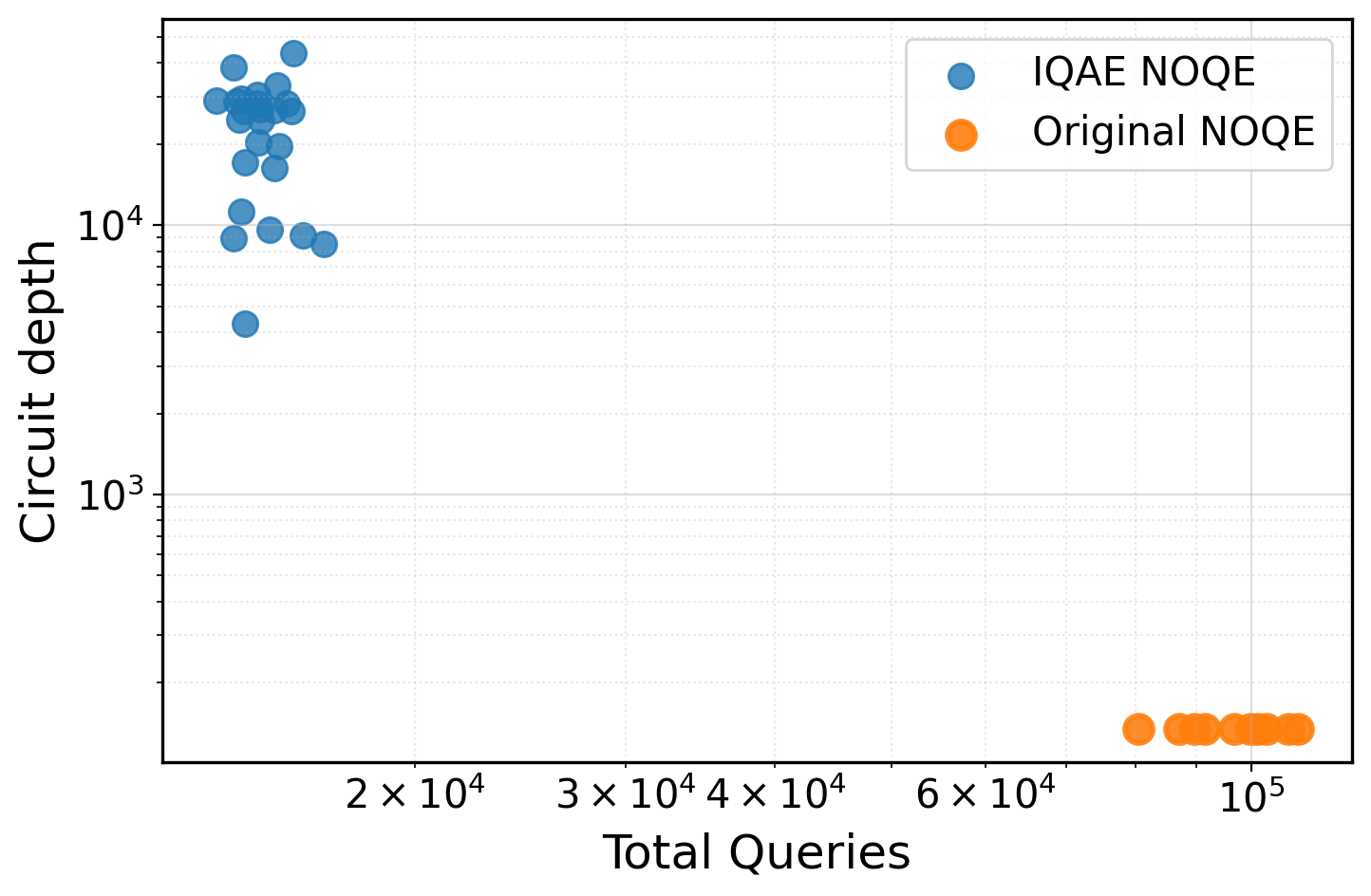}
    \caption{
    Tradeoff between total query count and circuit depth for IQAE--NOQE and the original NOQE.
    Each point corresponds to an independent energy-estimation instance.
    IQAE--NOQE substantially reduces the total number of queries by leveraging iterative amplitude estimation,
    at the cost of increased circuit depth due to repeated applications of the Grover operator.
    }
    \label{fig:comparison}
\end{figure}

\textbf{Resource Analysis and Tradeoffs}. \\
While IQAE--NOQE substantially reduces the total number of queries, this
improvement is achieved at the cost of increased circuit depth. To make this
tradeoff explicit, Fig.~\ref{fig:comparison} compares the total query count and circuit depth required
by IQAE--NOQE and the original NOQE.

Each point in Fig.~\ref{fig:comparison} corresponds to an independent
trial. The orange points show that the original NOQE operates with relatively shallow
circuits but requires a large number of queries due to repeated sampling. In
contrast, the blue points for IQAE--NOQE show a reduced query count by
amplifying the relevant amplitudes through repeated applications of the Grover
operator. This results in deeper circuits, as reflected by the increased circuit
depth along the vertical axis.

It is important to recognize that the tradeoff observed in Fig.~\ref{fig:comparison} is structural rather than incidental.
IQAE--NOQE concentrates resources into a small number of deeper circuits,
whereas the original NOQE distributes resources across a large number of shallow
measurements. Which approach is preferable therefore depends on the dominant
hardware constraints. For hardware and operating regimes where query cost is the primary resource bottleneck, the reduction in total queries provided by IQAE--NOQE can outweigh
the increased circuit depth. Conversely, in settings with strict coherence-time
limitations, the original NOQE measurement scheme may remain favorable. This explicit tradeoff
highlights the complementary regimes of applicability of the two protocols.

Taken together, the results in this section show that IQAE--NOQE provides a
systematic and quantitative improvement over sampling-based NOQE in measurement cost. The numerical comparisons
demonstrate that the reduction in query complexity predicted by amplitude
estimation directly translates into faster convergence of both matrix elements
and energies for a given precision. At the same time, the accompanying increase in
circuit depth is explicit and controllable, arising solely from repeated Grover
iterations. This clear separation of statistical and coherent resources makes
IQAE--NOQE a flexible alternative to the original NOQE, and motivates a broader
discussion of its practical implications and future extensions.

\section{Summary and Outlook}\label{sec:summary}

In this work, we revisited the non-orthogonal quantum eigensolver (NOQE) from the
perspective of measurement efficiency. The sampling-based estimation of Hamiltonian and overlap matrix elements leads to $\mathcal{O}(1/\varepsilon^2)$ scaling in measurement cost. By reformulating these matrix-element estimation tasks within the
framework of iterative quantum amplitude estimation, we introduced IQAE--NOQE,
which replaces incoherent statistical averaging with coherent information
accumulation prior to measurement and thereby improves the measurement cost to $\tilde{\mathcal{O}}(1/\varepsilon)$.

Numerical simulations for the hydrogen molecule demonstrate that IQAE--NOQE achieves accurate energy
estimation with approximately an order of magnitude fewer queries than the original NOQE measurement scheme using the Hadamard test. The results further show that the improved measurement efficiency
comes with an explicit tradeoff: a reduction in total query count at the cost of
increased circuit depth due to repeated Grover iterations. Importantly, this
tradeoff is structured and predictable, with the dominant cost concentrated in a
small number of high-precision refinement steps.

Several directions for future work naturally follow from this study.
First, it is important to note that quantum amplitude estimation admits many
algorithmic variants \cite{aaronson2020quantum,fukuzawa2023modified,callison2022improved,giurgica2022low}, which differ in their optimization targets, such as query
complexity, shot complexity, or circuit depth. In this work, we focus on the
most fundamental and enabling step: reformulating the matrix-element estimation
in NOQE as an amplitude estimation problem. Once this reformulation is in place,
other variants of amplitude estimation can be incorporated in a modular manner.
For example, alternative iterative schedules or low-depth variants may further
optimize performance under specific hardware constraints. A systematic
comparison of different amplitude estimation variants within the NOQE framework
is therefore a natural extension of the present work.

Second, experimental demonstrations of IQAE--NOQE on quantum hardware represent
an important next step. While  the results presented here isolate the projection (often also referred to as quantum measurement or quantum sampling) noise to assess the
intrinsic measurement efficiency of the protocol, realistic implementations on current hardware
will require a careful treatment of gate noise. Incorporating
error-mitigation strategies \cite{RevModPhys.95.045005,ren2024error} and noise-aware variants of amplitude estimation
will be essential for evaluating the practical viability of the approach on
near-term devices.

Finally, NOQE serves as a representative example of a broader class of quantum
algorithms whose performance is limited by sampling-based estimation of matrix
elements or expectation values. Other subspace-based and non-variational
algorithms, such as quantum subspace expansion methods or related projection-based
quantum algorithms \cite{yoshioka2022generalized,lee2024sampling,epperly2022theory}, share a similar structure and may also be reformulated within an
amplitude estimation framework. From this perspective, the present work can be
viewed as a heuristic step toward identifying and exploiting coherent
measurement reformulations more broadly, with the goal of inspiring similar
developments for other quantum algorithms where measurement cost is a dominant
bottleneck.

Overall, this work demonstrates that substantial improvements in quantum
algorithm performance can be achieved by revisiting how measurements are
formulated and executed, rather than by modifying state preparation or
 ans\"atze. By explicitly trading incoherent sampling for coherent
information accumulation, IQAE--NOQE provides a concrete example of how
measurement-limited quantum algorithms can be systematically transformed into
coherence-limited ones. We expect that this perspective will prove useful beyond
the specific setting studied here, and will help guide the development of
measurement-efficient quantum algorithms in regimes where both precision and
measurement cost are the dominant constraints, and allow analysis of the tradeoff between these two criteria.

\section{Acknowledgements}
 
The authors thank Martin Head-Gordon,
Nikolay Tkachenko, Mingyu Kang, Wendy Billings, Rebecca Tomann, Brandon Schramm,
Ayush Pancholy, Avijit Shee, and Alex Krotz for valuable discussions.

This work was partially supported by a joint development agreement between UC Berkeley and Dow, by the National Science Foundation (NSF) Quantum Leap Challenge Institutes (QLCI) program through Grant No. OMA-2016245, and by the U.S. Department of Energy, Office of Science, National Quantum Information Science Research Centers, Quantum Systems Accelerator. 

\section{Data Availability Statement}
The data that support the findings of this article are openly available \cite{ren2025measurementefficientdata}.

\clearpage
\newpage

\renewcommand{\thesection}{\Alph{section}}
\renewcommand{\thesubsection}{\thesection.\arabic{subsection}}

\titleformat{\section}
  [block]
  {\normalfont\Large\bfseries\centering}
  {Appendix \thesection:}
  {0.5em}
  {}

\titleformat{\subsection}
  [block]
  {\normalfont\large\bfseries\centering}
  {Appendix \thesection.\thesubsection:}
  {0.5em}
  {}

\appendix

\onecolumngrid

\addcontentsline{toc}{section}{APPENDIX}

\section*{Appendix: NOQE Algorithm and Implementation}
\label{appendix:noqe}

We briefly review the original non-orthogonal quantum eigensolver (NOQE)
algorithm and its circuit implementation. The purpose of this appendix is to
make explicit the sampling-based matrix-element estimation procedure used in
the original NOQE framework, which is replaced by IQAE-based estimation in the
main text.

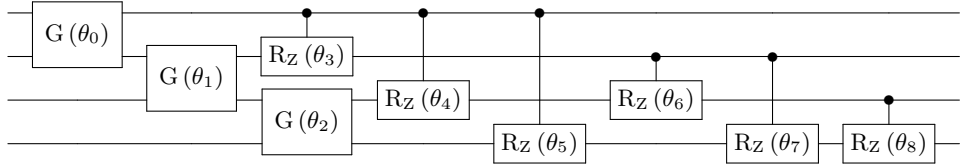
\begin{figure*}[!htbp]
    \centering
    \scalebox{1}{
    \Qcircuit @C=1.0em @R=0.2em @!R { \\
        \nghost{  } & \lstick{  } & \multigate{1}{\mathrm{G}\,(\mathrm{\theta_0})} & \qw & \ctrl{1} & \ctrl{2} & \ctrl{3} & \qw & \qw & \qw & \qw\\
        \nghost{ } & \lstick{  } & \ghost{\mathrm{G}\,(\mathrm{\theta_1})} & \multigate{1}{\mathrm{G}\,(\mathrm{\theta_1})} & \gate{\mathrm{R_Z}\,(\mathrm{\theta_3})} & \qw & \qw & \ctrl{1} & \ctrl{2} & \qw & \qw \\
        \nghost{} & \lstick{} & \qw & \ghost{\mathrm{G}\,(\mathrm{\theta_2})} & \multigate{1}{\mathrm{G}\,(\mathrm{\theta_2})} & \gate{\mathrm{R_Z}\,(\mathrm{\theta_4})} & \qw & \gate{\mathrm{R_Z}\,(\mathrm{\theta_6})} & \qw & \ctrl{1} & \qw \\
        \nghost{} & \lstick{} & \qw & \qw & \ghost{\mathrm{G}\,(\mathrm{\theta_6})} & \qw & \gate{\mathrm{R_Z}\,(\mathrm{\theta_5})} & \qw & \gate{\mathrm{R_Z}\,(\mathrm{\theta_7})} & \gate{\mathrm{R_Z}\,(\mathrm{\theta_8})} & \qw  \\
\\ }}
    \caption{Basic circuit subroutine for implementation of the UCCD operator $e^{\hat{\tau}}$ for H$_2$ in the STO-3G basis. $G(\theta)$ are Givens rotations that realize the $\tilde{\mathcal{U}}_{B}^{(l,\mu)}$ basis-rotation terms in Eq.~\eqref{lowrank}. The controlled-$Z$ rotations realize the $\hat{n}_{p}\hat{n}_{q}$ number-operator terms. Four consecutive subroutines of this form realize a single $e^{\hat{\tau}}$ ansatz for H$_2$ in the STO-3G basis. Single-qubit gates are omitted here for simplicity.}
    \label{fig:e_hat}
\end{figure*}

We use the unitary coupled-cluster doubles (UCCD) ansatz to prepare the
ansatz state:
\begin{equation}
    \left|\psi_{\mathrm{UCCD}}\right\rangle=e^{\hat{\tau}}\left|\psi_{\mathrm{UHF}}\right\rangle .
\end{equation}
Here $|\psi_{\mathrm{UHF}}\rangle$ is the unrestricted Hartree--Fock (UHF)
state, and $\hat{\tau}=\hat{T}-\hat{T}^{\dagger}$ with
\begin{equation}
    \hat{T}=\sum_{p q r s=1}^{N} t_{p s, q r}
    \hat{a}_{p}^{\dagger}\hat{a}_{q}^{\dagger}\hat{a}_{r}\hat{a}_{s}.
\end{equation}
Here $r,s\in \operatorname{\text{occ}}$ and $p,q\in \operatorname{virt}$. The
standard UCCD amplitudes $t_{ps,qr}$ can be approximated by second-order
many-body perturbation theory (MP2). More specifically, the UCCD operator
$e^{\hat{\tau}}$ can be decomposed into a sum of squares of normal operators
using a singular value decomposition:
\begin{equation}
    e^{\hat{\tau}}=\exp \left(-i \sum_{l=1}^{L} \sum_{\mu=1}^{m} \hat{Y}_{l,\mu}{ }^{2}\right).
\end{equation}
The normal operators are then Trotterized to obtain a low-rank representation
\cite{motta2021low}
\begin{equation}
\label{lowrank}
    e^{\hat{\tau}} \approx
    \hat{\mathcal{U}}_{B}^{(1,1)\dagger}
    \prod_{l=1}^{L}
    \prod_{\mu=1}^{m}
    \exp \left(
        -i \sum_{pq}^{\rho_l}
        \lambda_{p}^{(l,\mu)}
        \lambda_{q}^{(l,\mu)}
        \hat{n}_{p}\hat{n}_{q}
    \right)
    \tilde{\mathcal{U}}_{B}^{(l,\mu)} .
\end{equation}
Here $\lambda_{p}^{(l,\mu)}$ are eigenvalues of the $\hat{Y}_{l,\mu}$ operators,
and $\tilde{\mathcal{U}}_{B}^{(l,\mu)}$ are sequences of neighboring basis
rotations. The circuit implementation of the ansatz is illustrated in
Fig.~\ref{fig:e_hat}. Details of the derivation can be found in the original
NOQE paper \cite{baek2023say}.

\begin{figure}[htbp]
    \centering
    \includegraphics[width=0.6\linewidth]{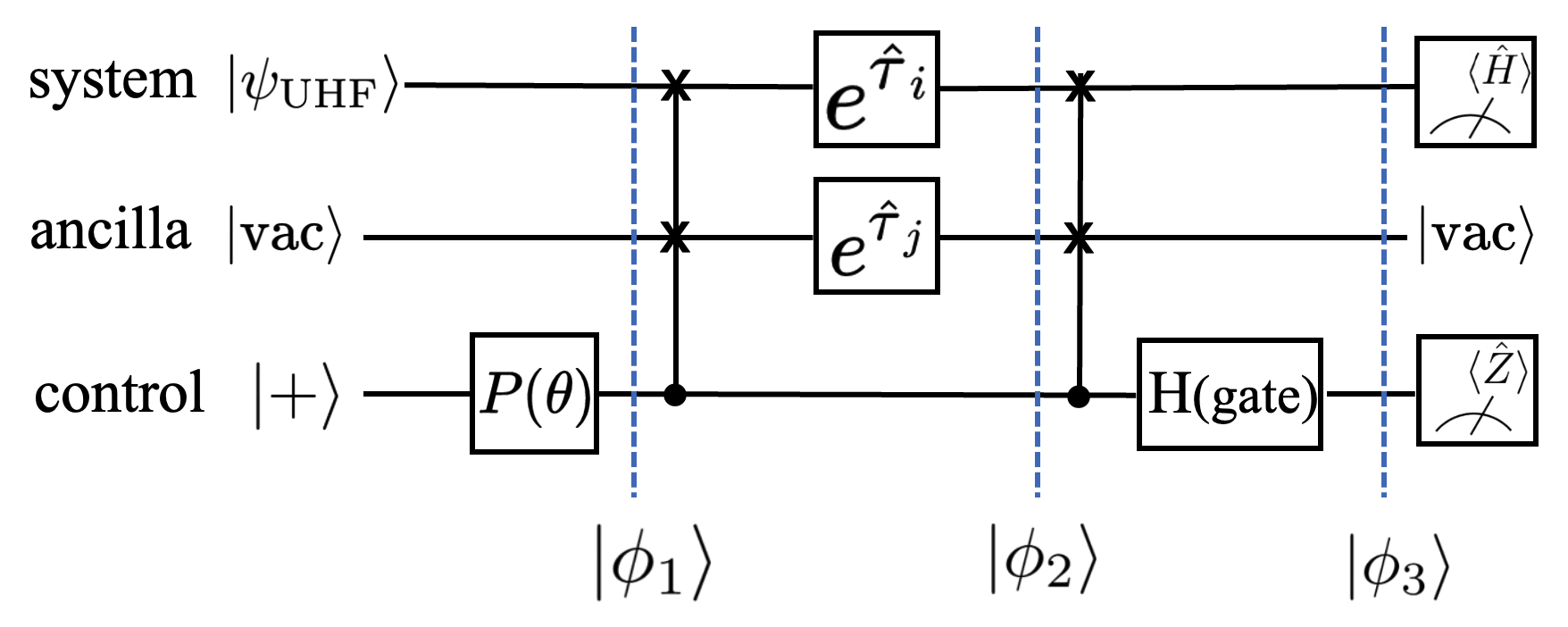}
    \caption{Original NOQE circuit. The quantum states at different stages of the circuit are marked and written in Eqs.~\eqref{sec:eq1}, \eqref{sec:eq2}, and \eqref{sec:eq3}.}
    \label{fig:noqe_circuit_appendix}
\end{figure}

The NOQE circuit for evaluation of elements of the Hamiltonian and overlap
matrices is given in Fig.~\ref{fig:noqe_circuit_appendix}. In this work, we
absorb the unitary orbital rotation operators $\hat{U}_{i\rightarrow 1}$ and
$\hat{U}_{j\rightarrow 1}$, which are required to transform $|\psi_i\rangle$
and $|\psi_j\rangle$ into the same single-particle basis, into the
corresponding $e^{\hat{\tau}_i}$ and $e^{\hat{\tau}_j}$ circuits.

The circuit starts with system qubits in the UHF state, ancilla qubits in the
all-zero state, and an extra control qubit in the $|+\rangle$ state. To begin,
a phase gate with angle $\theta$ is applied to the control qubit, creating
state $|\phi_1\rangle$:
\begin{equation}
\label{sec:eq1}
    \left|\phi_1\right\rangle
    =
    \frac{1}{\sqrt{2}}
    \left(
        \left|\psi_{\mathrm{UHF}}\right\rangle|\mathrm{vac}\rangle|0\rangle
        +
        e^{i\theta}
        \left|\psi_{\mathrm{UHF}}\right\rangle|\mathrm{vac}\rangle|1\rangle
    \right).
\end{equation}
An $N$-qubit controlled-swap (CSWAP) gate is then applied to transfer the
system state to the ancilla register, and distinct UCCD operators
$e^{\hat{\tau}_i}$ and $e^{\hat{\tau}_j}$ are applied to the system and ancilla
states. After absorbing the basis rotations into these ansatz operators, the
resulting state is
\begin{equation}
\label{sec:eq2}
    \left|\phi_2\right\rangle
    =
    \frac{1}{\sqrt{2}}
    \left(
        \left|\psi_i\right\rangle|\mathrm{vac}\rangle|0\rangle
        +
        e^{i\theta}
        |\mathrm{vac}\rangle\left|\psi_j\right\rangle|1\rangle
    \right),
\end{equation}
where the ansatz states $|\psi_i\rangle$ and $|\psi_j\rangle$ are now
expressed in the common basis. A second CSWAP conditioned on the control qubit
returns the ancilla register to $|\mathrm{vac}\rangle$. From now on, the
ancilla register can be ignored, and the final state is written in terms of the
system and control qubit:
\begin{equation}
\label{sec:eq3}
    \left|\phi_3\right\rangle
    =
    \frac{1}{\sqrt{2}}
    \left(
        \left|\psi_i\right\rangle|+\rangle
        +
        e^{i\theta}
        \left|\psi_j\right\rangle|-\rangle
    \right).
\end{equation}

We now address the measurement of Hamiltonian and overlap matrix elements.
Setting the angle $\theta$ of the phase gate to zero and measuring the control
qubit in the $Z$ basis results in the following post-measurement states in the
system register:
\begin{equation}
\label{post_meas_state}
\begin{aligned}
    \frac{\left|\psi_i\right\rangle+\left|\psi_j\right\rangle}
    {\sqrt{2+2 \operatorname{Re}\langle\psi_i \mid \psi_j\rangle}}
    \quad \quad
    \text{if $Z=+1$, with probability} \quad
    P(Z=+1)=\frac{1+\operatorname{Re}\langle\psi_i \mid \psi_j\rangle}{2};
    \\
    \frac{\left|\psi_i\right\rangle-\left|\psi_j\right\rangle}
    {\sqrt{2-2 \operatorname{Re}\langle\psi_i \mid \psi_j\rangle}}
    \quad \quad
    \text{if $Z=-1$, with probability} \quad
    P(Z=-1)=\frac{1-\operatorname{Re}\langle\psi_i \mid \psi_j\rangle}{2}.
\end{aligned}
\end{equation}
This implies that the measurement statistics of Pauli $Z$ on the control qubit
give the real part of the overlap matrix element,
\begin{equation}
    \langle \hat{I}\otimes \hat{Z}\rangle
    =
    \operatorname{Re}\langle\psi_i \mid \psi_j\rangle .
\end{equation}

To estimate the Hamiltonian matrix element $H_{ij}$, the Hamiltonian is first
decomposed into a sum of Pauli terms,
\begin{equation}
    \hat{H}
    =
    \sum_k \omega_k \hat{P}_k .
\end{equation}
After the measurement of the control qubit in the $\hat{Z}$ basis, each Pauli
operator $\hat{P}_k$ is measured on the system register. Measurements of
$\hat{P}_k$ on the post-measurement states in Eq.~\eqref{post_meas_state} give
\begin{equation}
\begin{aligned}
    \langle \hat{P}_k \rangle^{+}
    &=
    \frac{
        \langle \psi_i| \hat{P}_k|\psi_i\rangle
        +
        \langle \psi_j|\hat{P}_k|\psi_j\rangle
        +
        2 \operatorname{Re}\langle\psi_i| \hat{P}_k|\psi_j\rangle
    }
    {
        2+2 \operatorname{Re}\langle\psi_i \mid \psi_j\rangle
    },
    \\
    \langle \hat{P}_k \rangle^{-}
    &=
    \frac{
        \langle \psi_i| \hat{P}_k|\psi_i\rangle
        +
        \langle \psi_j|\hat{P}_k|\psi_j\rangle
        -
        2 \operatorname{Re}\langle\psi_i| \hat{P}_k|\psi_j\rangle
    }
    {
        2-2 \operatorname{Re}\langle\psi_i \mid \psi_j\rangle
    }.
\end{aligned}
\end{equation}
Combining these conditional expectation values with their corresponding
probabilities gives
\begin{equation}
    \operatorname{Re}\langle\psi_i| \hat{P}_k|\psi_j\rangle
    =
    P(Z=+1)\langle \hat{P}_k \rangle^{+}
    -
    P(Z=-1)\langle \hat{P}_k \rangle^{-}.
\end{equation}
Equivalently, this quantity can be obtained by measuring the joint observable
$\hat{P}_k\otimes \hat{Z}$.

Summing over the Pauli decomposition gives the real part of the Hamiltonian
matrix element,
\begin{equation}
    \operatorname{Re}H_{ij}
    =
    \sum_k
    \omega_k
    \operatorname{Re}\langle\psi_i| \hat{P}_k|\psi_j\rangle .
\end{equation}
By setting $\theta=\pi/2$ and repeating the same procedure, one obtains the
imaginary parts of the overlap and Hamiltonian matrix elements,
$\operatorname{Im}\langle\psi_i|\psi_j\rangle$ and
$\operatorname{Im}\langle\psi_i|\hat{H}|\psi_j\rangle$, up to the sign
convention set by the phase gate.

Once all matrix elements have been estimated,
\begin{equation}
\label{hs_matrix}
    H_{ij}
    =
    \langle \psi_i|\hat{H}|\psi_j\rangle,
    \qquad
    S_{ij}
    =
    \langle \psi_i|\psi_j\rangle ,
\end{equation}
the electronic energies are obtained by solving the generalized eigenvalue
problem
\begin{equation}
\label{generalized_eigenvalue_problem}
    \mathbf{H}\vec{c}
    =
    E\mathbf{S}\vec{c}
\end{equation}
on a classical computer. In the original NOQE protocol, the matrix elements are
obtained by repeated projective measurements. Therefore, estimating each matrix
element to additive precision $\varepsilon$ requires
$O(1/\varepsilon^2)$ circuit repetitions at fixed confidence. This
sampling-based step is the part replaced by IQAE in the main text.


%

\end{document}